\documentclass{article}

\def\mytitle#1{\setcounter{equation}{0}
\setcounter{footnote}{0}
\begin{flushleft}\Large\textbf{#1}\end{flushleft}
\vspace{0.25cm}}
\def\myname#1{\leftline{{\large #1}}\vspace{-0.13cm}}
\def\myplace#1#2{\small\begin{flushleft}\textit{#1}\\
\texttt{#2}\end{flushleft}}
\newenvironment{contribution}{\normalsize\noindent}{}
\def\myclassification#1{\small\noindent
 PACS No : 
  #1\vspace{0.5cm}}
\usepackage{graphicx}
\begin{document}

\mytitle{Accretion of Chaplygin gas upon black holes: Formation of faster outflowing winds}

\myname{ Ritabrata Biswas~$*$}
\vskip0.2cm
\myname{ Subenoy Chakraborty~$*$}
\vskip0.2cm
\myname{ Tarun Deep Saini~$\dag$}
\vskip0.2cm
\myname{ Banibrata Mukhopadhyay~$\dag$}

\myplace{$*$Department of Mathematics, Jadavpur University, Kolkata-70032, India.}
{[biswas.ritabrata@gmail.com, schakraborty@math.jdvu.ac.in]}


\myplace{$\dag$ Department of Physics, Indian Institute of Science, Bangalore-560012, India.}
{[tarun@physics.iisc.ernet.in, bm@physics.iisc.ernet.in]}

\begin{abstract}
We study the accretion of modified Chaplygin gas upon different types of black hole. Modified Chaplygin gas is one of the best candidates for 
a combined model of dark matter and dark energy. In addition, from a field theoretical point of view the modified Chaplygin gas model 
is equivalent to that of a scalar field having a self-interacting potential. We formulate the equations related to 
both spherical accretion and disc accretion, and respective
winds. The corresponding numerical solutions of the flow, particularly of velocity, are 
presented and are analyzed. We show that the accretion-wind system of modified Chaplygin gas dramatically alters the wind solutions, producing
faster winds, upon changes 
in physical parameters, while accretion solutions qualitatively remain unaffected. This implies that modified Chaplygin gas is more prone to
produce outflow which is the natural consequence of the dark energy into the system. \\

\end{abstract}

\myclassification{98.62.Mw, 95.36.+x, 98.62.Mw, 95.30.Sf}

\begin{contribution}
\end{contribution}
\section{Introduction}
In nature, the compact objects, particularly the black holes (BHs), are not visible but can be detected by the presence of the accretion disc around them. By analyzing  light rays off an accretion disc, one can speculate the properties of the central compact object. The formation of accretion discs can most commonly be understood in a binary system where accretion of matter into a compact object from the companion star forms a disc like structure. The proto-planetary discs, discs around active galactic nuclei etc are also examples of accretion disc \cite{Mukhopadhyay1}.

Although the accretion phenomena around compact objects (particularly BHs) have been extensively discussed over the last three decades (e.g. \cite{Mukhopadhyay1}), it was started long ago in 1952 by Bondi  \cite{Bondi}. He studied stationary spherical accretion problem by introducing formal fluid dynamical equations in the Newtonian framework. In the framework of general relativity, the study of accretion was initiated by Michel \cite{Michel}. By choosing the Newtonian gravitational potential, Shakura and Sunyaev \cite{Shakura} formulated very simplistic but effective model of the accretion disc. Some aspects of the accretion disc in fully relativistic framework had been studied by Novikov and Thorne \cite{Novikov} and Page and Thorne \cite{Page}. Subsequently, various aspects related to the critical behavior of general relativistic flows in spherical symmetry have been studied \cite{Begelman, Brinkmann, Malec, Das}. Although there are a few steps forward, still it is extremely difficult to simulate the full scale realistic accretion discs including outflows in a full general relativistic framework.

 One may note that in BH accretion, an important issue is that the flow of accreting matter must be transonic in nature, i.e., there should be sonic point(s) \cite{Chakraborti1, Chakraborti2, Chakraborti3} in the flow. On the other hand, accretion flow around a neutron star is not necessarily transonic (i.e., sonic point may or may not exist). 
It depends on the inner disc boundary conditions influenced by the neutron star. Note that, 
to simplify the nonlinearity arised due to general relativity, often pseudo-Newtonian approach is used by introducing 
the pseudo-Newtonian potential for the accretion disc. In this scheme, to study the accretion flow, the dynamical equations are written in the Newtonian theory with an effective gravitational force corresponding to the above pseudo-Newtonian potential. Paczynski and Wiita (PW) \cite{Paczynski}  first proposed such a pseudo-Newtonian potential for a non-rotating BH and it has been frequently used in simulations \cite{Milsom, Hawley}. Subsequently, there were several other proposals \cite{Mukhopadhyay2, Chakraborti4} for pseudo-Newtonian potentials, but it has been shown \cite{Artemova} that the former one (by PW) is better than the others for non-rotating compact objects. Then  Mukhopadhyay \cite{Mukhopadhyay3} proposed a pseudo-Newtonian potential for rotating compact objects directly from the space-time metric. He was able to reproduce exactly the general relativistic values of last stable circular orbits and had shown in calculating specific energies of last stable circular orbits that the difference between the general relativistic value  and that corresponding to his potential was less than 10$\%$ in Kerr geometry. Also when BH spin is set to zero, the potential of Mukhopadhyay \cite{Mukhopadhyay3} reduces to that of PW. 

At present, there are various observed data (particularly
from distant type-Ia supernova explosions) which strongly indicate
that our universe is undergoing an acceleration phase starting at
the cosmological red-shift $z\simeq 1$ \cite{Spergel1, Spergel2}. Now the reason for this
accelerating expansion can be described in two ways : - either by
introducing non-baryonic matter (known as dark energy (DE)) having negative pressure (violating
strong energy condition $\rho+3p<0$) in the framework of general
relativity or in the framework of modified gravity theory, such as
$f(R)$ gravity, where the extra terms in the modified Friedmann
equations are responsible for the acceleration. Also these recent observational results
support various cosmological tests such as gravitational lensing,
galaxy number counts etc. \cite{Ostriker}

Within the framework of Einstein's gravity, due to this present accelerating phase, it is reasonable to believe  that DE is the dominating part ($74.5\%$ of the energy content in the observable universe) \cite{Bahcall} of the total energy of the universe. The candidates for DE \cite{Copeland, Alcaniz} can be classified as cosmological constant, dynamical component like quintessence, \cite{Wetterich}, k-essence \cite{Armendariz1}, Chaplygin gas (CG) \cite{Kamenhchik, Gorini} etc. Although the cosmological constant is by far the simplest and the most popular candidate for DE, from the point of view of fine tuning and cosmic coincidence problem, it is not a suitable candidate for DE. On the other hand, dynamical DE models are favorable as they admit to construct `tracker' \cite{Zlatev} or `attractor' \cite{Armendariz2} solutions. But most of the dynamical DE 
models are described by a scalar field (often called quintessence field) which is unable to describe the transition 
from a universe filled with matter to an exponentially expanding universe. However, the DE can be represented by 
an exotic type of fluid known as CG \cite{Kamenhchik} having equation of state $p=-\frac{\beta}{\rho}$, where 
$p$ and $\rho$ are respectively the pressure and energy density and $\beta$ is a positive constant. 
Subsequently, this equation of state was generalized to $p=-\frac{\beta}{\rho^{n}},~ 0\leq n \leq 1$, and is known as generalized CG (GCG) \cite{Gorini}. In recent past, there was further modification to this equation of state as \cite{Benaoum}
\begin{equation}\label{1}
p=\alpha \rho - \frac{\beta}{\rho^{n}}
\end{equation}
and the model is known as modified CG (MCG). It gives the cosmological evolution from an initial radiation era (with $\alpha=\frac{1}{3}$) to (asymptotically) the $\Lambda CDM$ era (where fluid behaves as cosmological constant). This cosmological model can be described from the field theoretical point of view by introducing a scalar field $\phi$ having self-interacting  potential $V(\phi)$ so that the effective Lagrangian 
$$L_{\phi}=\frac{1}{2}\dot{\phi}^{2}-V(\phi).$$
Moreover, the MCG model is interacting from phenomenological view point and it can be motivated by braneworld interpretation \cite{Bento}. Further, the MCG model is in favor of the cosmological tests namely gravitational lensing \cite{Silva}, gamma-ray bursts \cite{Bertolami} in addition to the above mentioned observational evidences. Unlike GCG, the velocity of sound with MCG is compatible with the cosmic evolution. Also the MCG model (as well as the GCG model) is naturally constrained through the cosmological observable \cite{Sandvik} to explain the overwhelming energy density of the universe at present era. At low energy density, the model behaves as polytropic gas \cite{Santos}  with negative value of the index and hence it is possible to have astrophysical implications of the models with an alternate way of restricting the parameters.

In the present work, we plan to study the flow of MCG around BHs. Due to present accelerated expansion of the universe, the matter in the universe is dominated by DE (almost $\frac{2}{3}$ of the matter is in the form of DE). Therefore it is reasonable to assume the accreting matter is in the form of DE. 
Babichev et.al. \cite{Babichev1, Babichev2} were the pioneers to think about the DE accretion upon a BH, in the framework
of Bondi accretion \cite{Bondi}. 
Further, as we have stated above, MCG is one of the most favorable candidates for DE filling our universe mostly.
However, as matter approaches the BH, the centrifugal force increases rapidly compared to the gravitational force (particularly for the flow with slowly
varying specific angular momentum). Hence, matter feels increasing centrifugal force around a radius little away from the BH, then slows down and piles up around the
radius. This region refers to CENBOL (CENtrifugal pressure supported BOundary Layer) (see, e.g., \cite{dc99}) acting as the effective boundary layer of the 
BH system which, like stellar surface, could produce outflowing winds. Therefore, due to the presence of CENBOL, a BH system, like star, easily exhibits
wind of matter along with inflow. However, once matter crosses the CENBOL, the dominant gravitational force solely controls dynamics and matter
falls into the BH.


The size and shape of galaxies depend strongly on their angular momentum; spiral galaxies containing discs possess
a net sense of rotation while elliptical galaxies, that are without discs, possess none. 
Hoyle \cite{Hoyle1} first proposed 
gravitational instability, arising from gravitational coupling with the surrounding matter (tidal interactions), as 
a possible explanation for galactic rotation. Alternatively, people have proposed \cite{vonWiezsacker1, Gamow1} that 
the origin of galaxy rotation could be due to primordial turbulence/vorticity. 
However, vortical modes decay with time and, therefore, a significant vorticity at the time of galaxy formation would 
typically require an unrealistic magnitude of vorticity in the early universe.
At present, it is widely believed that the hierarchical clustering of cold dark matter \cite{Blumenthal1} is the origin 
of structures in the universe. Consequently, the angular momentum of dark matter halos and eventually the rotation of 
galaxies is thought to be produced by gravitational tidal torque \cite{Barnes1}. It has been suggested that the halos 
obtain their spins through the cumulative acquisition of angular momentum from satellite accretion. DE that has accreted 
on a galaxy would be similarly torqued by such tidal interactions. Although it is difficult to estimate this 
effect without invoking specific models, it is reasonable to expect that some angular momentum might reside in 
the DE halos of galaxies. If such rotating DE were to accrete on a compact object then it would carry a part of 
its angular momentum with it, thus leading to the formation of a DE accretion disc as opposed to Bondi accretion.

Therefore, we have concentrated on MCG accretion-wind disc system around BHs, along with Bondi flow. 
We have examined the accretion and wind solutions extensively. The results are then compared with the standard 
accretion-wind phenomena with flow having polytropic equation of state. The summary of the work is presented at the end.

\section{Basic equations}
 Bondi \cite{Bondi} formulated simple spherical accretion and wind in the Newtonian framework around a non-rotating star. As gravity is very strong near compact objects, general relativistic  treatment gives proper description of the accretion flow. However, the full general relativistic set of equations is so cumbersome that it is difficult to relate different terms with the 
corresponding physics behind them and as a result transparency of the description might be hidden. Thus we describe the system by pseudo-Newtonian approach, i.e., Newtonian theory with a modified gravitational potential (force). 
This helps to describe the general relativistic features approximately and useful to deal with a less understood system.

For an isolated compact object, matter falls onto it from all directions, resulting in a spherical accretion as defined by Bondi \cite{Bondi}. 
However, as described in \S 1, infall onto a super-massive BH at the center of a spiral galaxy forms an accretion disc.
Here we consider the inviscid fluid so that there is no dissipative force into the system and then the specific 
angular momentum of disc fluid is constant throughout a particular flow. This importantly implies that 
accretion of MCG onto a black hole may be hindered for the same reasons that the accretion
of standard cold dark matter is hindered \cite{pdef,hu}.
Therefore, in steady state the radial momentum conservation equation describing disc dynamics is \cite{Mukhopadhyay1}
\begin{equation}\label{2}
u\frac{du}{dx}+\frac{1}{\rho}\frac{dp}{dx}-\frac{\lambda^{2}}{x^{3}}+F_{g}(x)=0,
\end{equation}
where all variables are expressed in dimensionless units as follows:\\
$u=u_{r}=\frac{v}{c},~x=\frac{r}{r_{g}},~r_{g}=\frac{GM}{c^{2}}$, where $c_{s}$, $M$ and $c$ are the speed of sound, mass of the BH and the speed of light respectively and $r$, $v$ are dimensionful radial coordinate and radial velocity respectively.
Here $p$ and $\rho$ are the dimensionless isotropic pressure and density of the flow, $F_{g}=\frac{\left(x^{2}-2 j\sqrt{x}+j^{2}\right)^{2}}{x^{3}\left(\sqrt{x}(x-2)+j\right)^{2}}$, is the gravitational force corresponding 
to the pseudo-Newtonian potential  \cite{Mukhopadhyay3} and $\lambda$ is the dimensionless specific angular momentum 
(in units of $\frac{GM}{c}$) of the flow, $j$ is the dimensionless specific angular momentum of the BH (Kerr parameter). In the
case of Bondi flow $\lambda=0$ throughout. The equation of continuity, i.e., vertically integrated mass conservation relation,
in general for disc accretion, gives
\begin{equation}\label{3}
\frac{d}{dx}\left(x u \Sigma\right)=0,
\end{equation}
where
$\Sigma$, the vertically integrated density, is given by \cite{Santos}
\begin{equation}\label{4}
\Sigma~~=~~I_{c}\rho_{e}h(x),
\end{equation}
when,\\
$I_{c}$  =  constant (related to  the  equation of state of  the accreting fluid) \cite{Mukhopadhyay1},\\
$\rho_{e}$ = density  at  equatorial  plane,\\
$h(x)$  = half-thickness  of  the  disc.\\
Assuming the vertical equilibrium, the expression for $h(x)$ can be written as:
\begin{equation}\label{5}
h(x)=c_{s}\sqrt{\frac{x}{F_{g}}}
\end{equation}
where $c_{s}^{2}= \frac{\partial p}{\partial \rho}\sim\frac{p}{\rho}$ is the square of sound speed.
Note that for spherical accretion equation (\ref{3}) reduces to 
$$
\frac{d}{dx}(x^2 u \rho)=0.
$$ 

In the present work, the accreting matter is chosen to be MCG. From the equation of state (given by equation (\ref{1})) 
of MCG, we obtain \\

\begin{equation}\label{5a}
c_{s}^{2}=\frac{\partial p}{\partial \rho}=\alpha+\frac{\beta n}{\rho^{n+1}}
\label{ex1}
\end{equation}
such that
\begin{equation}\label{6}
\rho~~=~~\left(\frac{n \beta}{c_{s}^{2}-\alpha}\right)^{\frac{1}{(n+1)}}
\end{equation}
and thus
$$\frac{1}{\rho}\frac{dp}{dx}=-\frac{2c_{s}^{3}}{\left(n+1\right)\left(c_{s}^{2}-\alpha\right)}\frac{dc_{s}}{dx}$$
\begin{equation}\label{6a}
=-\frac{1}{n+1}\frac{d}{dx}\left(c_{s}^{2}\right)-\left(\frac{\alpha}{n+1}\right)\frac{d}{dx}\left\{\ln\left(c_{s}^{2}-\alpha\right)\right\}.
\end{equation}
Now integrating equations (\ref{2}) and (\ref{3}) and using the MCG equation of state (\ref{1}) we obtain two conservation equations namely,
(a) energy conservation : -
\begin{equation}\label{7}
\varepsilon=\frac{1}{2}u^{2}-\frac{c_{s}^{2}}{n+1}-\frac{\alpha}{n+1}\ln \left(c_{s}^{2}-\alpha\right)+\frac{1}{2}\frac{\lambda^{2}}{x^{2}}+V(x),
\end{equation}
where $V(x)=\int F_{g}(x) dx$,

(b) conservation of mass : -
\begin{equation}\label{8}
\dot{M}=\odot \rho c_{s} \frac{x^{\frac{3}{2}}}{F_{g}^{\frac{1}{2}}}u=\odot c_{s}u\left(\frac{x^{3}}{F_{g}}\right)^{\frac{1}{2}}\left(\frac{n \beta}{c_{s}^{2}-\alpha}\right)^{\left(\frac{1}{n+1}\right)},
\end{equation}
where $\odot$ is a geometric constant, depending on the exact geometry of the flow. The entropy accretion/wind rate ($\dot{\mu}$) is related to the mass accretion/wind rate ($\dot{M}$), given by $\dot{\mu}=K^{n}\dot{M}$ ($K$ is a constant carrying 
information of entropy) so that \cite{Mukhopadhyay1}
\begin{equation}\label{9}
\dot{\mu}\simeq \odot c_{s} u \left(\frac{x^{3}}{F_{g}}\right)^{\frac{1}{2}}\left(\frac{n \beta}{c_{s}^{2}-\alpha}\right)^{\left(\frac{1}{1+n}\right)}.
\end{equation}
Now differentiating the $\rho$ replaced part of equation (\ref{8}) we obtain
\begin{equation}\label{9c}
\frac{\left(1-n\right)c_{s}^{2}+\alpha\left(n+1\right)}{\left(n+1\right)c_{s}\left(c_{s}^{2}-\alpha\right)}\frac{dc_{s}}{dx}=\frac{3}{2x}-\frac{1}{2F}\frac{dF}{dx}+\frac{1}{u}\frac{du}{dx}
\end{equation}
and then combining equations (\ref{6a}) and (\ref{9c}) to replace $\frac{1}{\rho}\frac{dp}{dx}$ in equation (\ref{2}) we obtain
\begin{equation}\label{10}
\frac{du}{dx}=\frac{\frac{\lambda^{2}}{x^{3}}-F_{g}(x)+\left(\frac{3}{x}-\frac{1}{F_{g}}\frac{dF_{g}}{dx}\right)\frac{c_{s}^{4}}{\left\{\left(1-n\right)c_{s}^{2}+\alpha\left(n+1\right)\right\}}}{u-\frac{2c_{s}^{4}}{u\left\{\left(1-n\right)c_{s}^{2}+\alpha\left(n+1\right)\right\}}}=\frac{N}{D}.
\end{equation}

Equation (\ref{10}) has two branches of solution, accretion (inflow) and wind (outflow).
Note that, for accretion, far away from BH, $u<c_{s}$ and very close to it $u>c_{s}$, while for wind it is opposite. However, in either of the cases,
there is an intermediate location where the denominator of equation (\ref{10}) vanishes. Thus, for a smooth solution throughout, the numerator has to be zero at that location. This point/location is called the sonic point or critical point ($x_{c}$). In studying accretion-wind phenomena, the existence of critical point plays an important role. Further, critical point always exists in accretion disc around BHs and its global analysis helps to understand the stability of physical parameters.
Now as $N=0=D$ at the critical point ($x=x_{c}$), using l'Hospital's rule and after some algebra, the velocity gradient of the accreting matter at the critical point obeys the quadratic equation
\begin{equation}\label{11}
A\left(\frac{du}{dx}\right)^{2}_{x=x_c}+B\left(\frac{du}{dx}\right)_{x=x_c}+C=0,
\end{equation}
where $$A=2\left[1-\frac{2\left(c_{sc}^{2}-\alpha\right)\left(n+1\right)\left\{\left(1-n\right)c_{sc}^{2}+2\alpha\left(n+1\right)\right\}}{\left\{\left(1-n\right)c_{sc}^{2}+\alpha\left(n+1\right)\right\}^{2}}\right],$$
\begin{equation}\label{12}
B=-\frac{2}{c_{sc}^{4}}\frac{\left(c_{sc}^{2}-\alpha\right)\left(n+1\right)\left\{\left(1-n\right)c_{sc}^{2}+2\alpha\left(n+1\right)\right\}}{\left\{\left(1-n\right)c_{sc}^{2}+\alpha\left(n+1\right)\right\}}\left[F_{g}(x_{c})-\frac{\lambda^{2}}{x_{c}^{3}}\right],
\end{equation}
$$C=\left(\frac{3\lambda^{2}}{x_{c}^{4}}-\left(\frac{dF_{g}}{dx}\right)_{x=x_{c}}\right)-\left\{\left(\frac{1}{F_{g}}\left(\frac{dF_{g}}{dx}\right)^{2}\right)_{x=x_{c}}-\frac{3}{x_{c}^{2}}-\left(\frac{1}{F_{g}}\frac{d^{2}F_{g}}{dx^{2}}\right)_{x=x_{c}}\right\}\frac{u_{c}^{2}}{2}~~~~~~~~~~~~~~~~~~~~~~~~~~~~~~
$$$$~~~~~~~~~~~~~~~~~~~~~~~~~~~~~~~~~~~~~~~~-\frac{u_{c}^{2}}{2c_{sc}^{8}}\left[\left(c_{sc}^{2}-\alpha\right)\left(n+1\right)\left\{\left(1-n\right)c_{sc}^{2}+2\alpha\left(n+1\right)\right\}\right]\left[F_{g}(x_{c})-\frac{\lambda^{2}}{x_{c}^{3}}\right]^{2},$$
where $c_{sc}$ denotes the speed of sound at the critical point.
Solving $N=0=D$ at the critical point, the Mach number and the flow velocity at that location, 
respectively denoting as $M_c$ and $u_c$, are given by
\begin{equation}\label{13}
M_{c}=\frac{u_{c}}{c_{sc}}=\left[\frac{2c_{sc}^{2}}{\left\{\left(1-n\right)c_{sc}^{2}+\alpha\left(1+n\right)\right\}}\right]^{\frac{1}{2}}
\end{equation}
and
\begin{equation}\label{14}
c_{sc}=\left[\frac{(1-n)\left(\frac{\lambda^{2}}{x_{c}^{3}}-F_{g}\left(x_{c}\right)\right)}{2\left\{\left(\frac{1}{F_{g}}\frac{dF_{g}}{dx}\right)_{x=x_{c}}-\frac{3}{x_{c}}\right\}}\left\{1+\sqrt{1+\frac{4\alpha(n+1)}{(1-n)^{2}}\frac{\left\{\left(\frac{1}{F_{g}}\frac{dF_{g}}{dx}\right)_{x=x_{c}}-\frac{3}{x_{c}}\right\}}{\left\{\frac{\lambda^{2}}{x_{c}^{3}}-F_{g}\left(x_{c}\right)\right\}}}\right\}\right]^{\frac{1}{2}}.
\end{equation}
Thus integrating the velocity gradient (i.e., equations (\ref{10}) and (\ref{11})) with appropriate boundary conditions 
\cite{Mukhopadhyay1} we can obtain the flow properties. The solution procedure has been discussed in detail in 
previous work \cite{cm99,rm10}, which is not repeated here.

It is to be noted that if we put $\alpha=0, ~\beta~\rightarrow~-\kappa,~n\rightarrow~-\gamma$, then we obtain $p=\kappa\rho^{\gamma}$.
In that case
the resulting differential equation of $u$, expressions for $A, B, C$ of equation (\ref{12}), Mach number, speed of 
sound, match with the Mukhopadhyay's results \cite{Mukhopadhyay1}.

\section{Analysis of solutions for spherical accretion-wind}
At the quintessence or phantom era the whole universe is supposed to be filled up with the DE fluid. In this case, 
it will be very natural that DE will fall upon a BH from every possible direction forming a spherical accretion-wind system. 
So obviously $\lambda$ will be zero.
\begin{figure}
(a)~~~~~~~~~~~~~~~~~~~~~~~~~~~~~~~~~~~~~~~~~~~~~~~~~~~~~~~~(b) \\ 
\includegraphics[height=2.6in, width=2.6in]{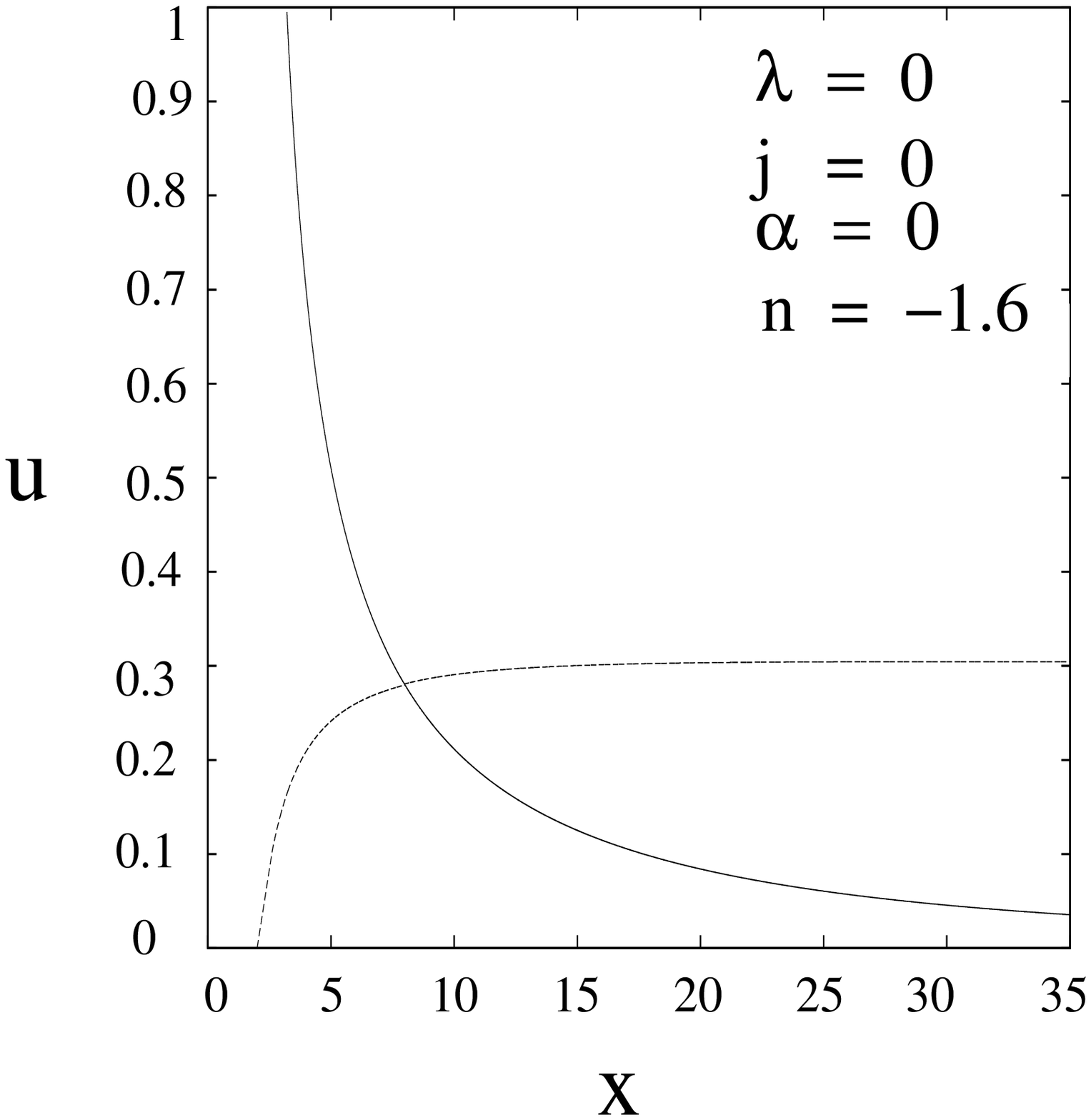}~~
\includegraphics[height=2.6in, width=2.6in]{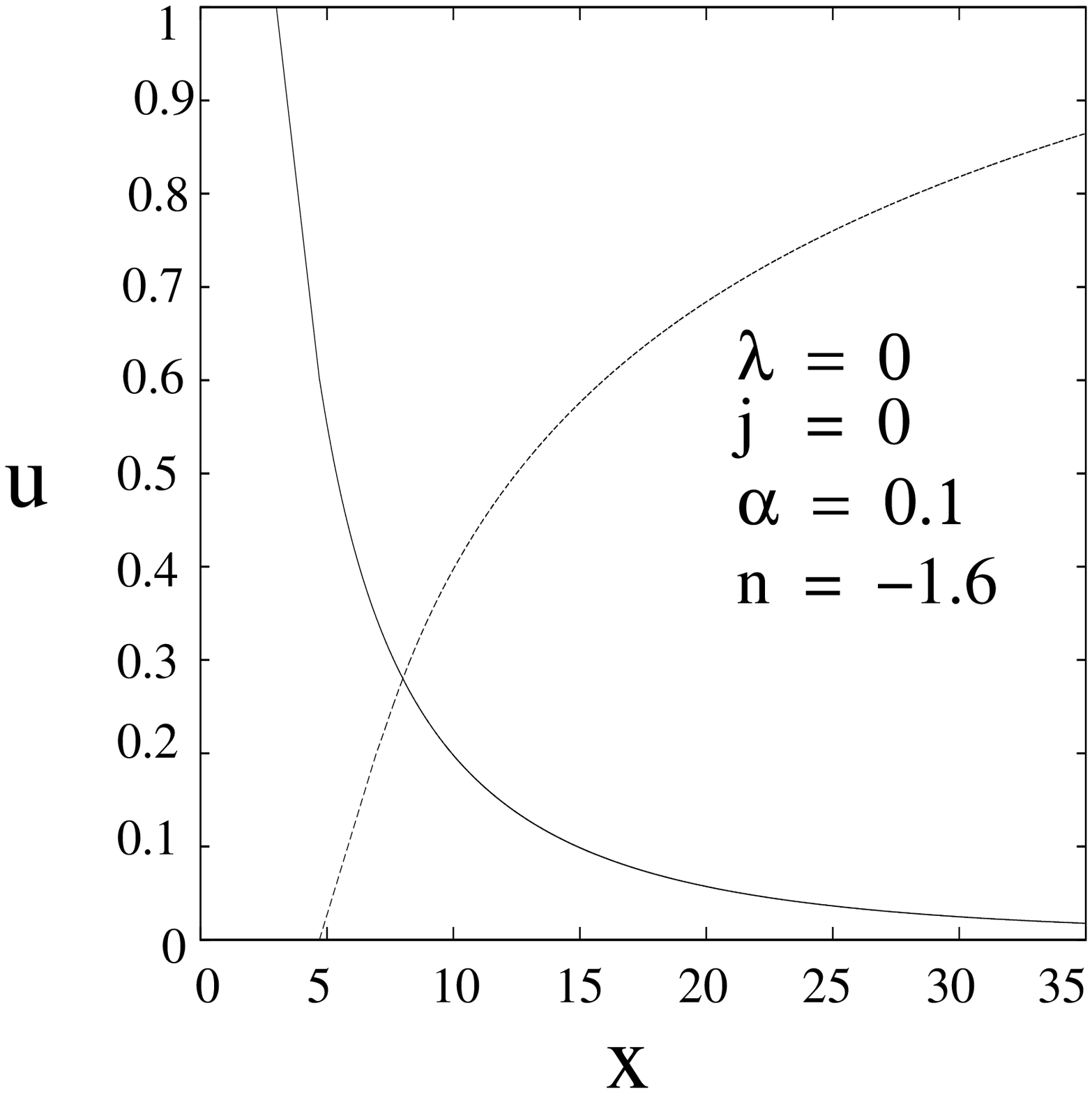}\\ \\
\hspace{1cm}
~~~~~~~~~~~~~~~~~~~~~~~~~~~~~~~~~~~~~~(c)~~~~~~~~~~~~~~~~~~~~~~~~~~~~~~~~~~~~~~~~~~~~~~~~~~~~~~~~(d) \\ 
\includegraphics[height=2.6in, width=2.6in]{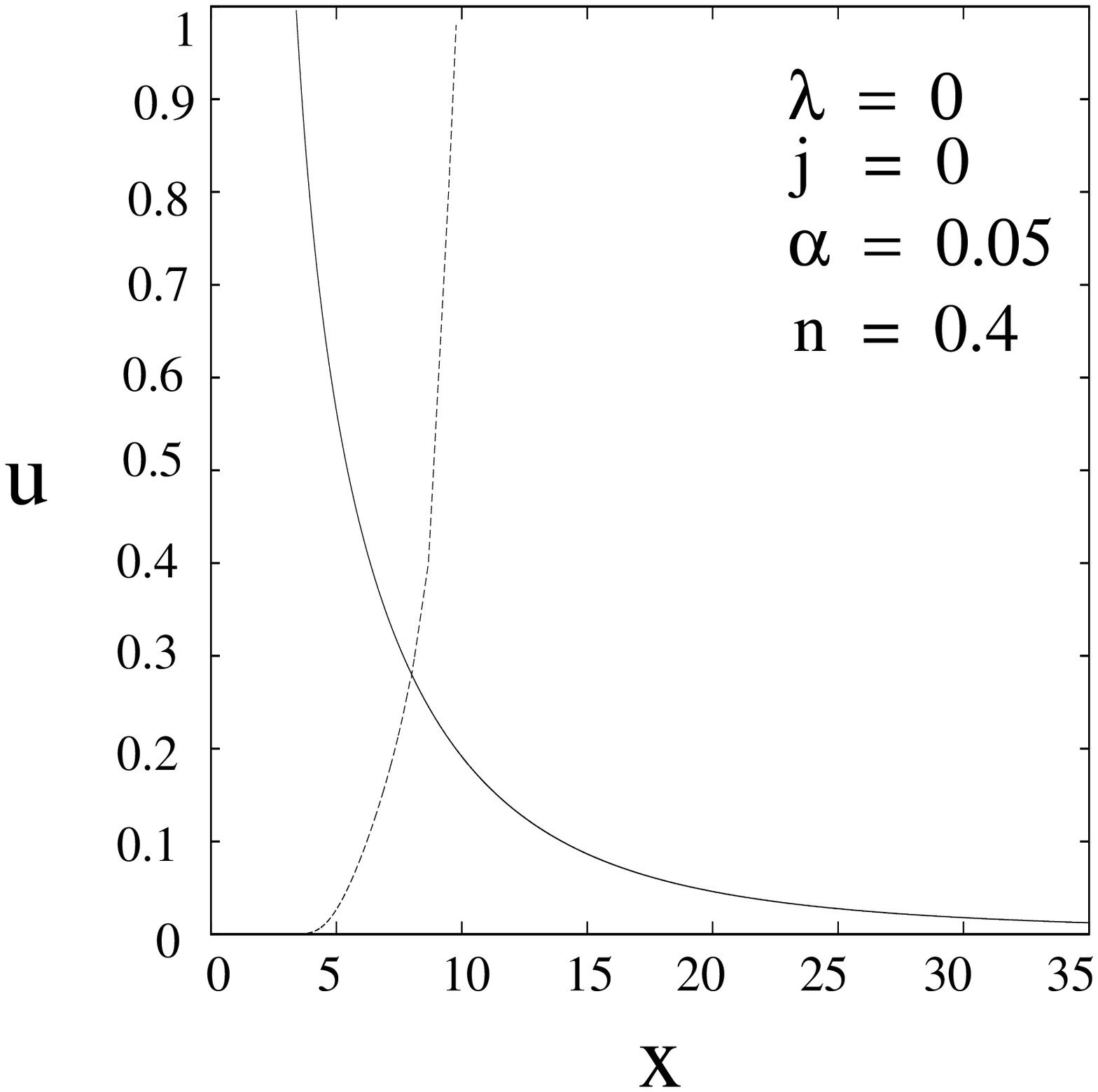}~~
\includegraphics[height=2.6in, width=2.6in]{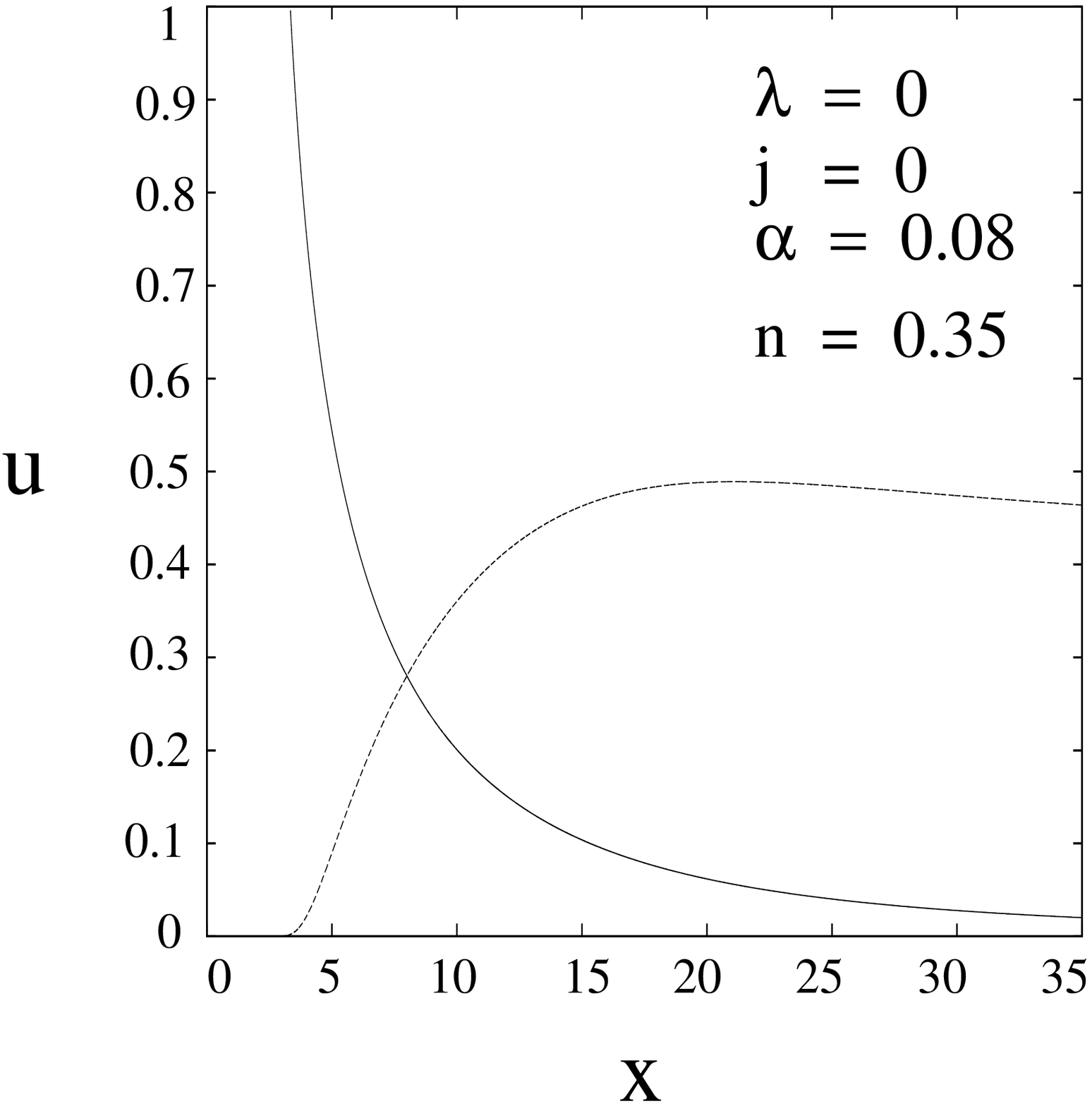}\\ \\
Fig. 1(a)-1(d) represent the variation of accretion and wind velocities as functions of radial coordinate for $\lambda=0$, $j=0$. The solid lines represent the accretion whereas the dotted lines are for wind. Note that the
magnitude of the velocities is plotted here. 
\hspace{1cm}
\vspace{2cm}
\end{figure}
In the figs 1(a)-(d) given here we have two distinct curves-- one represents the velocity of accreting matter and the other the 
velocity of the wind, i.e., the matter thrown outwards from the BH. As BH is a gravitationally attracting body, the velocity of accretion gradually 
increases as matter approaches to the BH and reaches almost to the speed of light near the BH event horizon. On the other hand, while 
the wind velocity is very small near the BH (it is almost impossible to escape out when matter is very near to BH), it increases 
gradually as matter comes away off the BH. The point of intersection of these velocity profiles, i.e., the radius where accretion 
velocity is equal to wind velocity, is same as critical point. Now the position of critical point depends upon many parameters. 
If matter is such that it has a tendency to easily increase its velocity, then the rate of change of velocity increases
faster for accretion
as matter approaches the BH and that for wind it decreases. Hence the critical point appears far away from the BH. 
On the other hand, 
if matter has a tendency to be flown outward, then the rate of change of velocity increases faster for wind
as matter comes away off
the BH and that for accretion decreases. As the wind velocity becomes equal to the accretion velocity at a nearer 
to the BH, critical 
point appears close to the BH, in the second case. 

We have studied various possible solutions for $j=0$. The ranges of parameters of MCG are chosen in order to obtain
physical solutions such that the accretion extends from infinity to the BH event horizon. 
When $\alpha=0$ and $n$ is negative, i.e., the adiabatic case, 
the solutions match exactly with that of the adiabatic Bondi flow, as shown in fig 1(a).
In fig 1(b), we have increased the value of $\alpha$ and noted that the wind rate increases significantly. 
Note that this is same as isothermal Bondi flow, as the first term in MCG equation of state dominates over the 
second one when $\rho<1$.
In figs 1(c) and 1(d) $n$ is increased and the increase of wind rate is revealed in fig 1(c) 
significantly. This is because at a high (positive) $n$, negative term in MCG in equation (\ref{1}) dominates significantly 
over the first term for $\rho<1$ rendering a very high negative pressure, which results in a faster wind. 
However, once the values of $\alpha$ and $n$ change to their highest (physically) possible values, as chosen 
in fig 1(d), $c_{sc}$ becomes smaller than the chosen $\sqrt{\alpha}$. From equation (\ref{6}) this gives a physical 
solution only if $\beta<0$. However, this does not correspond to a negative pressure rendering a slower wind in 
fig 1(d). This issue has been discussed again in \S 6 in order to discuss sound speed in the flows.

\section{Analysis of solutions for disc accretion-wind}

\begin{figure}
(a)~~~~~~~~~~~~~~~~~~~~~~~~~~~~~~~~~~~~~~~~~~~~~~~~~~~~~~~~(b) \\ 
\includegraphics[height=2.6in, width=2.6in]{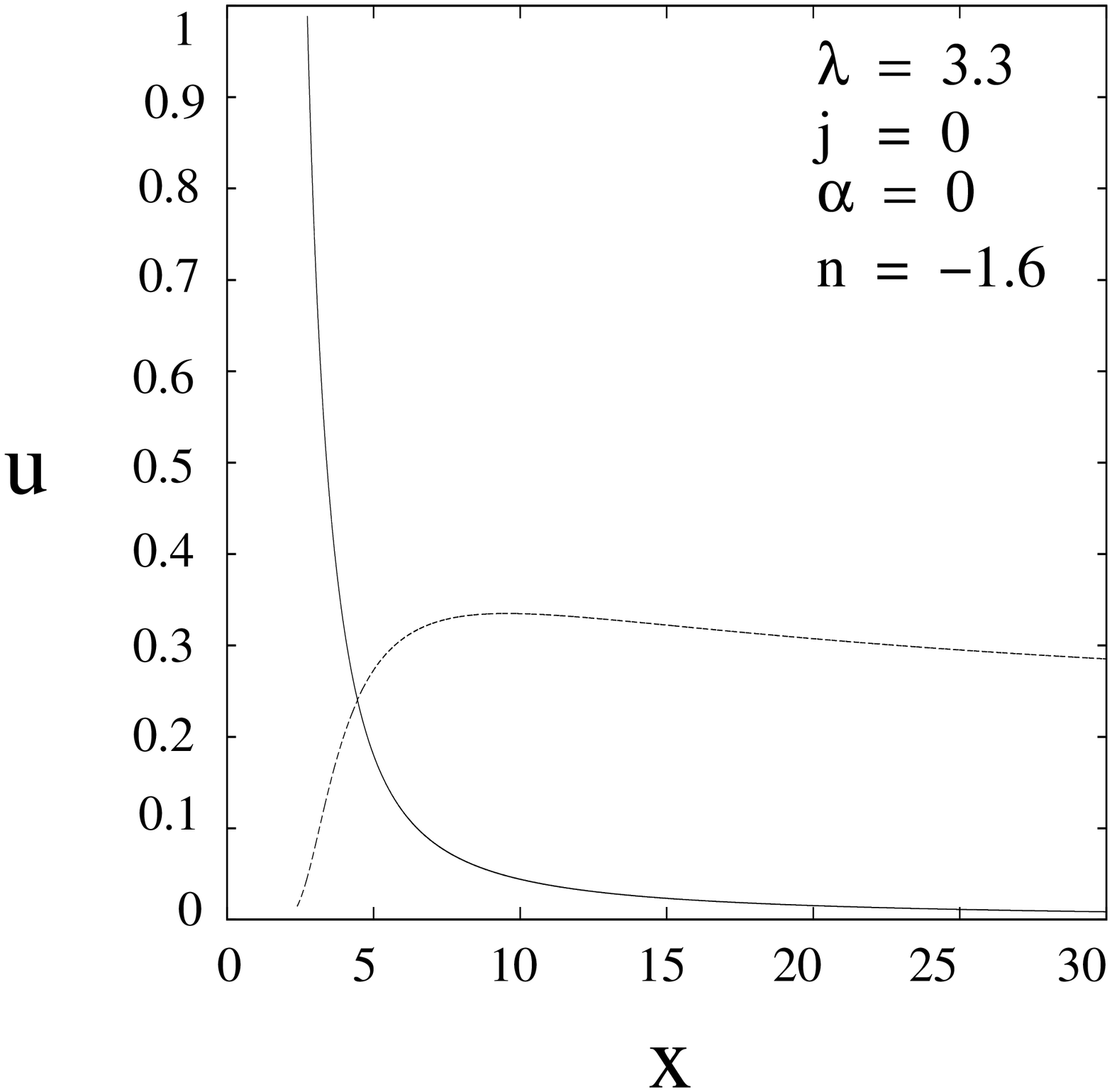}~~
\includegraphics[height=2.6in, width=2.6in]{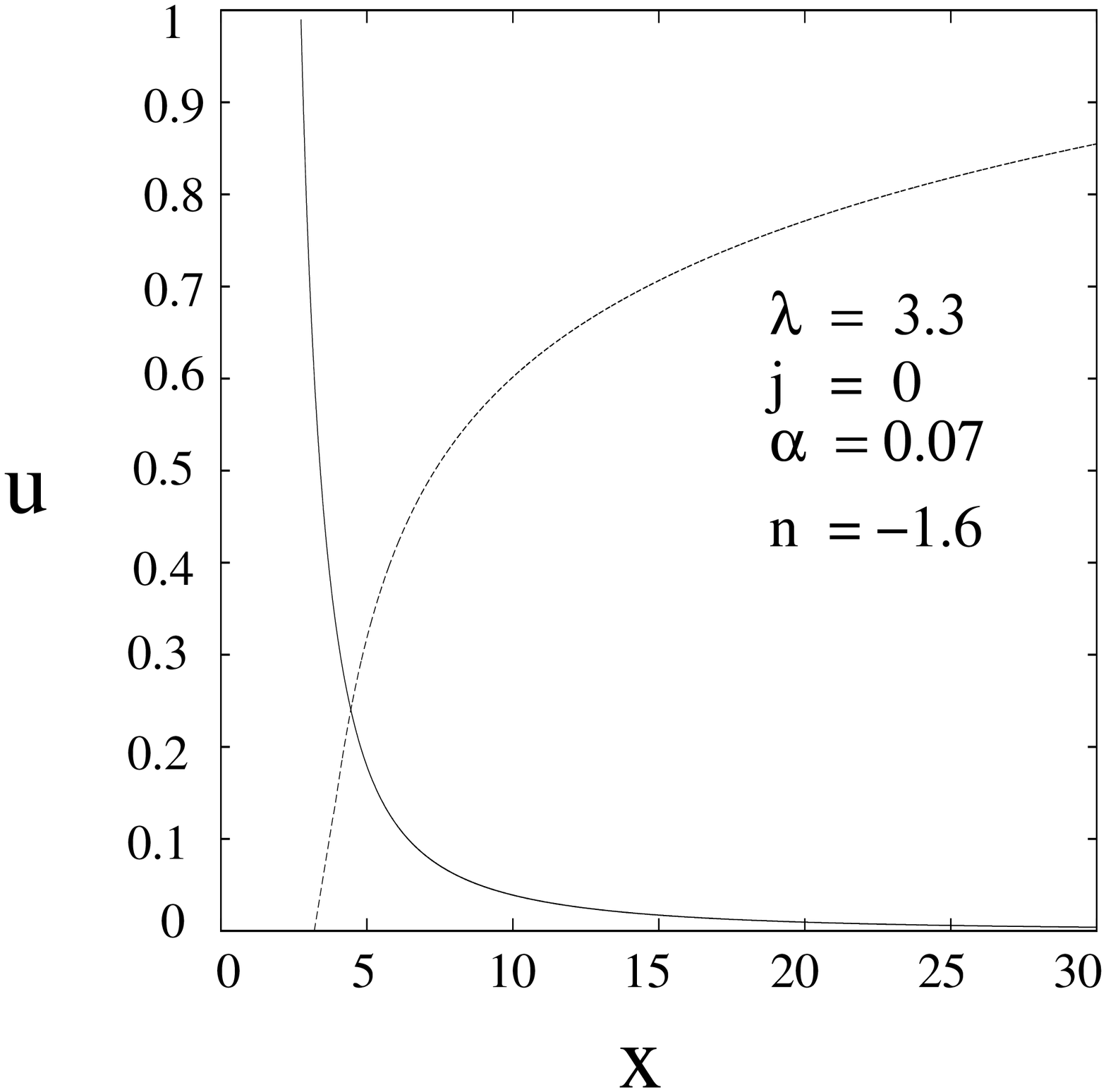}\\ \\
\hspace{1cm}
~~~~~~~~~~~~~~~~~~~~~~~~~~~~~~~~~~~~~~(c)~~~~~~~~~~~~~~~~~~~~~~~~~~~~~~~~~~~~~~~~~~~~~~~~~~~~~~~~(d) \\ 
\includegraphics[height=2.6in, width=2.6in]{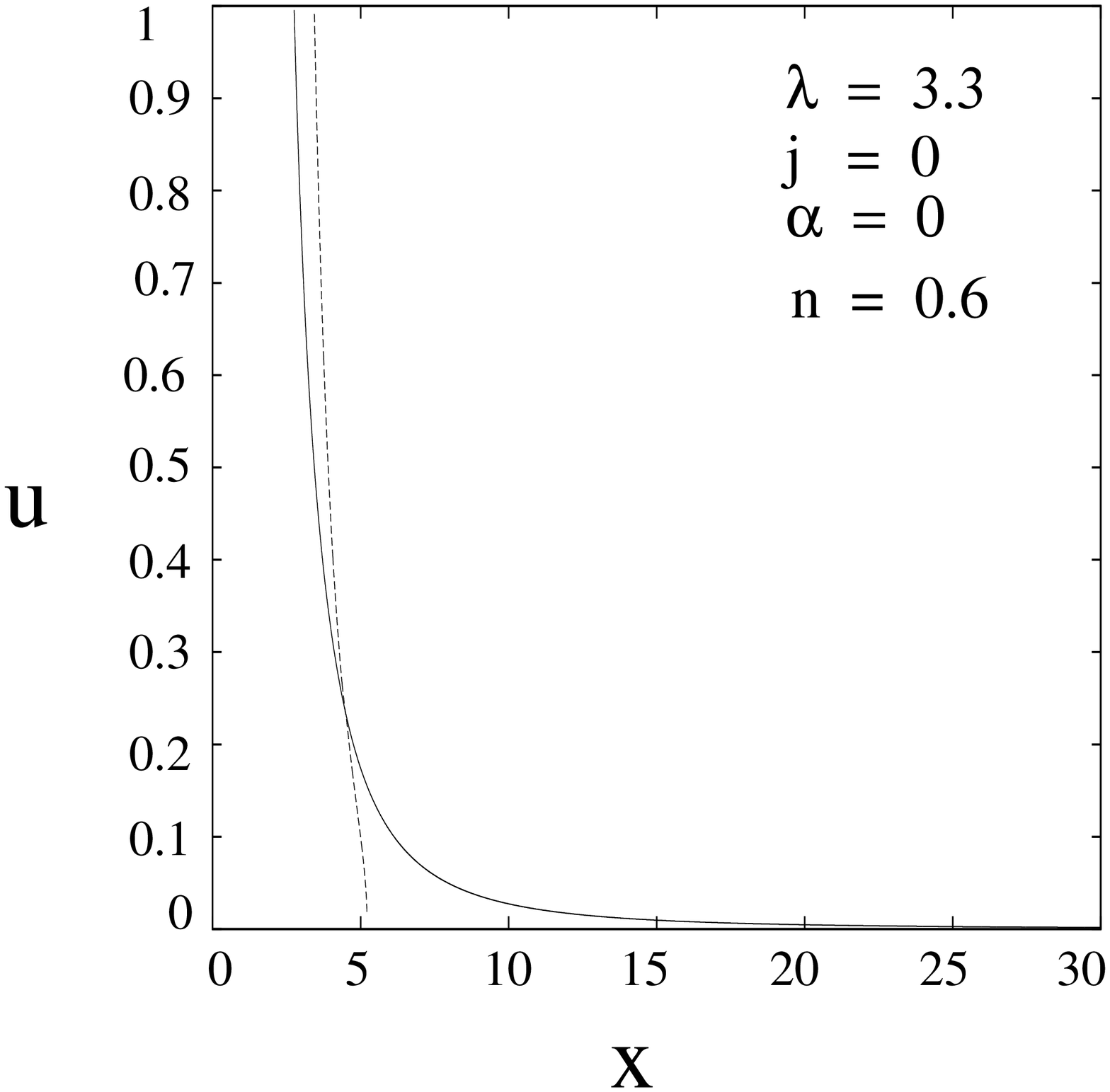}~~
\includegraphics[height=2.6in, width=2.6in]{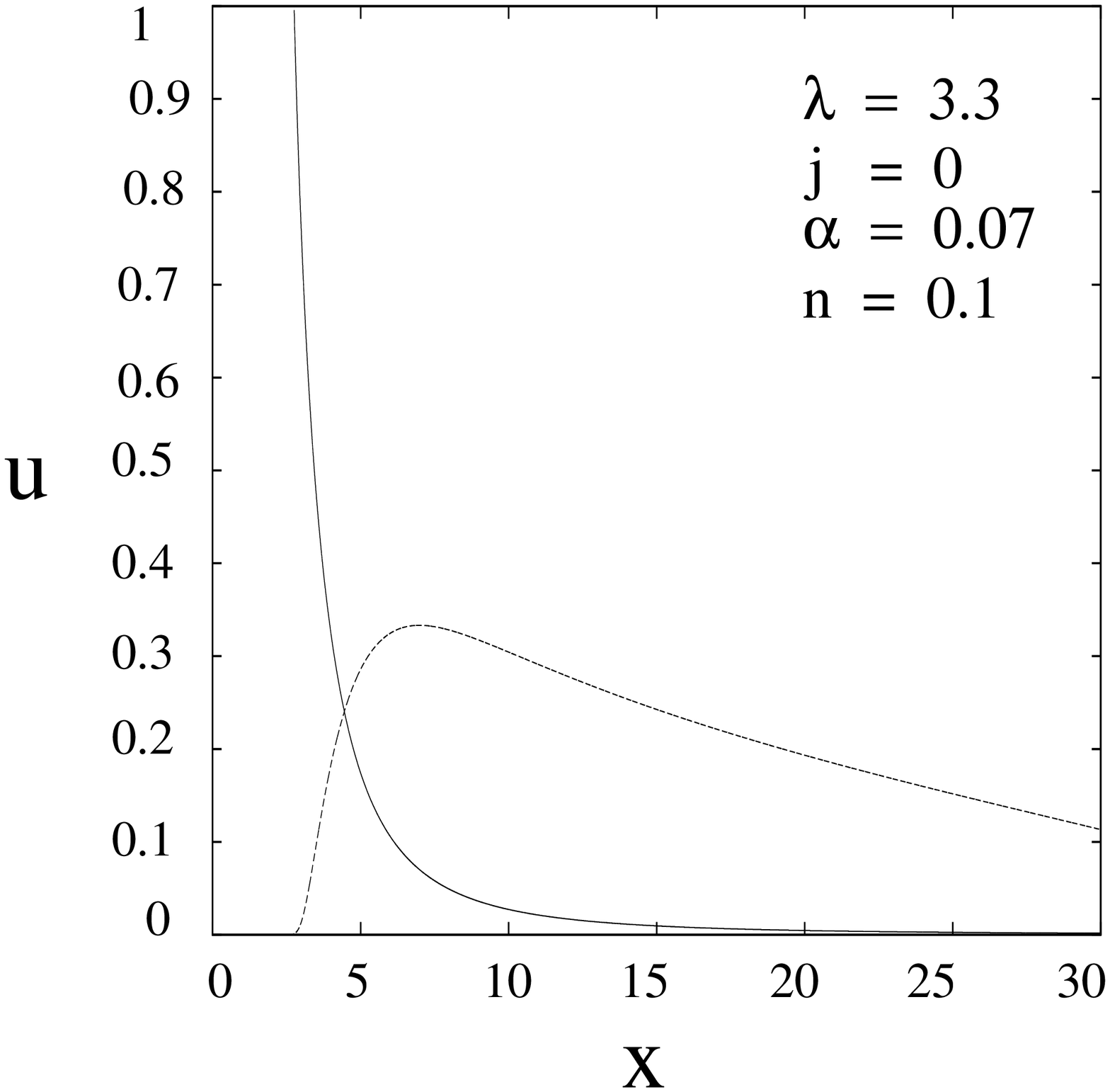}\\ \\
Fig. 2(a)-2(d) represent the variation of accretion and wind velocities in disc flows as functions of radial coordinate for $j=0$. The solid lines represent the accretion whereas the dotted lines are for wind. 
Note that the magnitude of velocities is plotted.
\hspace{1cm}
\vspace{2cm}

\end{figure}
\begin{figure}
(a)~~~~~~~~~~~~~~~~~~~~~~~~~~~~~~~~~~~~~~~~~~~~~~~~~~~~~~~~(b) \\ 
\includegraphics[height=2.6in, width=2.6in]{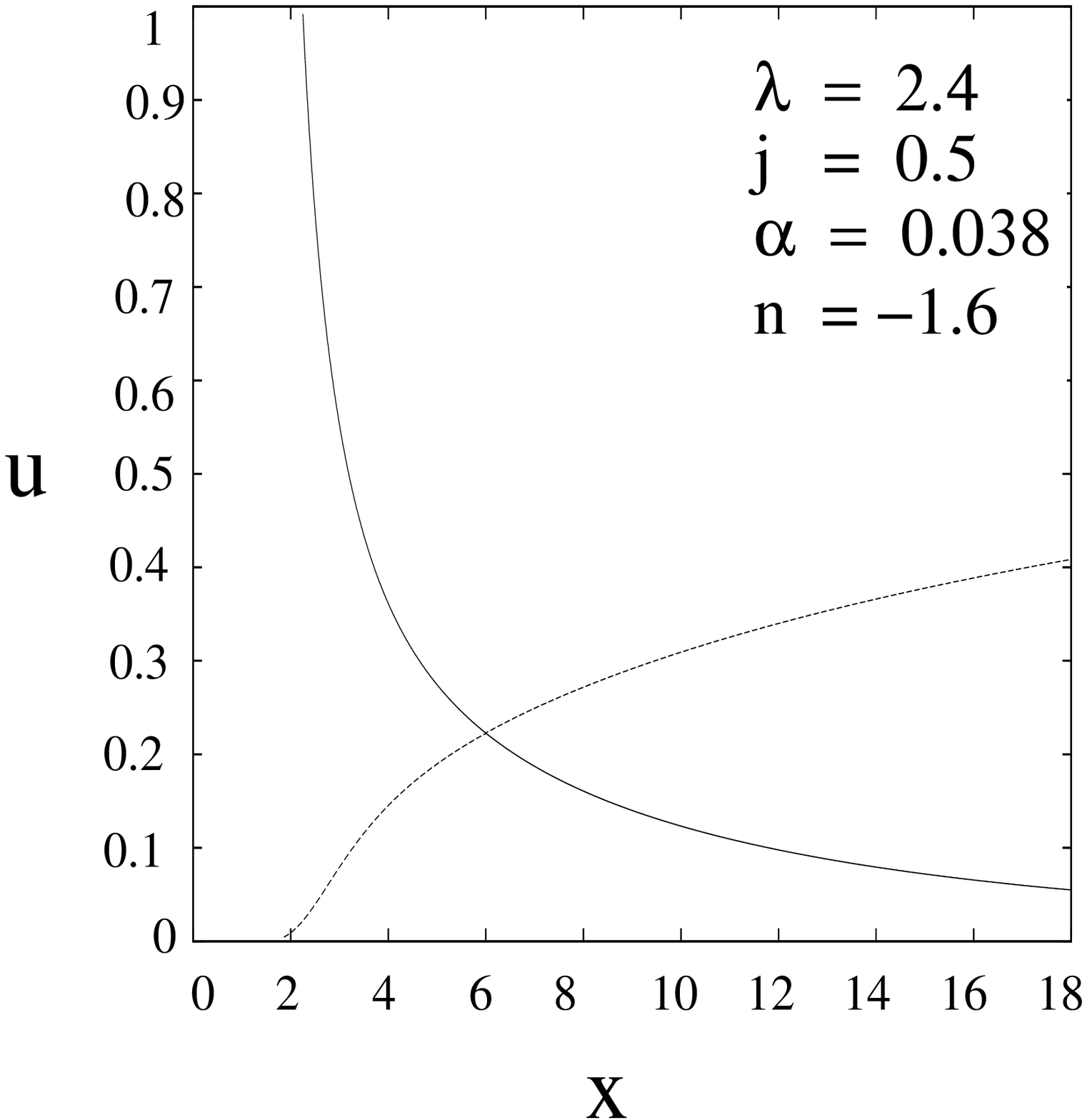}~~
\includegraphics[height=2.6in, width=2.6in]{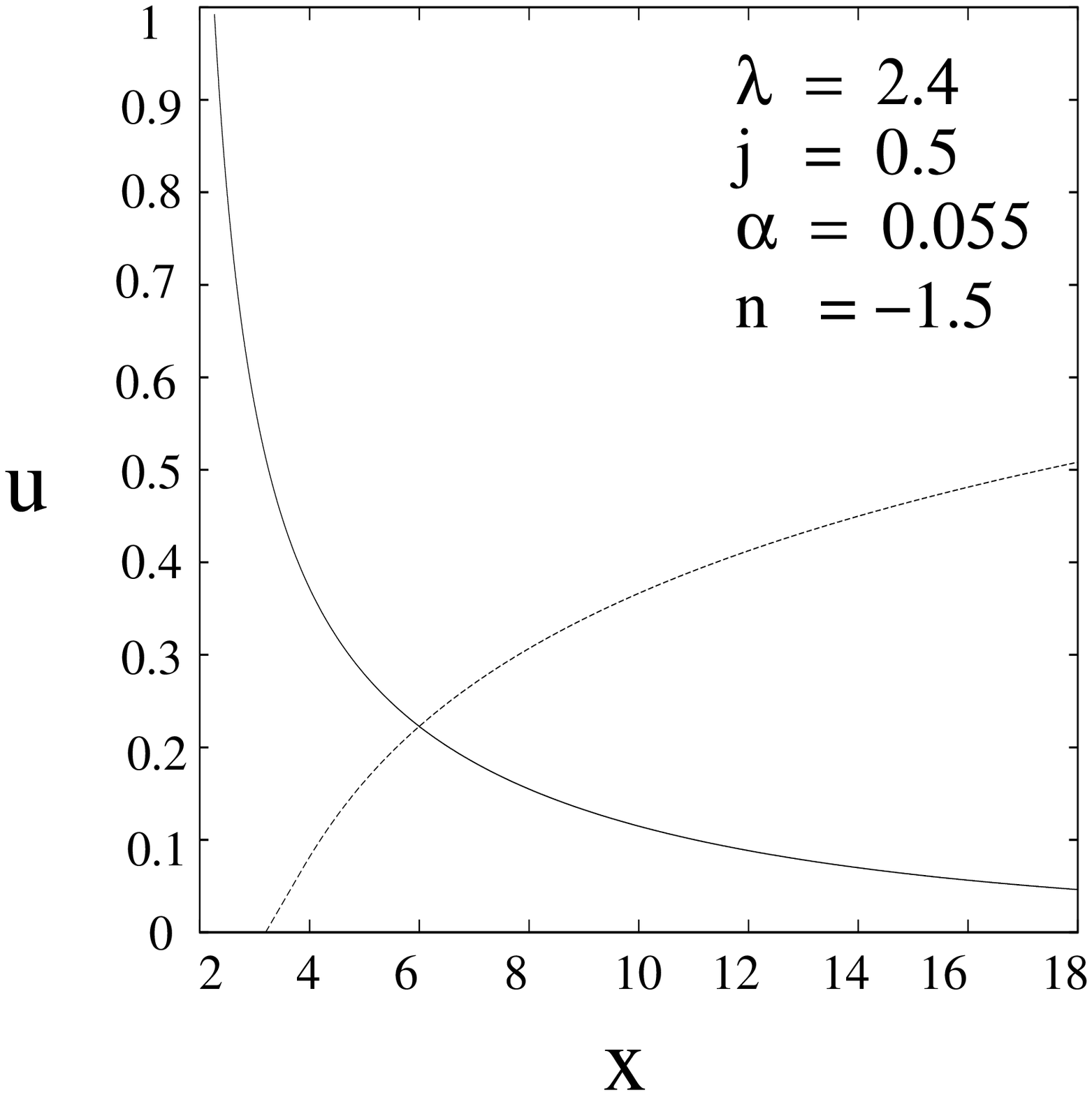}\\ \\
\hspace{1cm}
~~~~~~~~~~~~~~~~~~~~~~~~~~~~~~~~~~~~~~(c)~~~~~~~~~~~~~~~~~~~~~~~~~~~~~~~~~~~~~~~~~~~~~~~~~~~~~~~~(d) \\ 
\includegraphics[height=2.6in, width=2.6in]{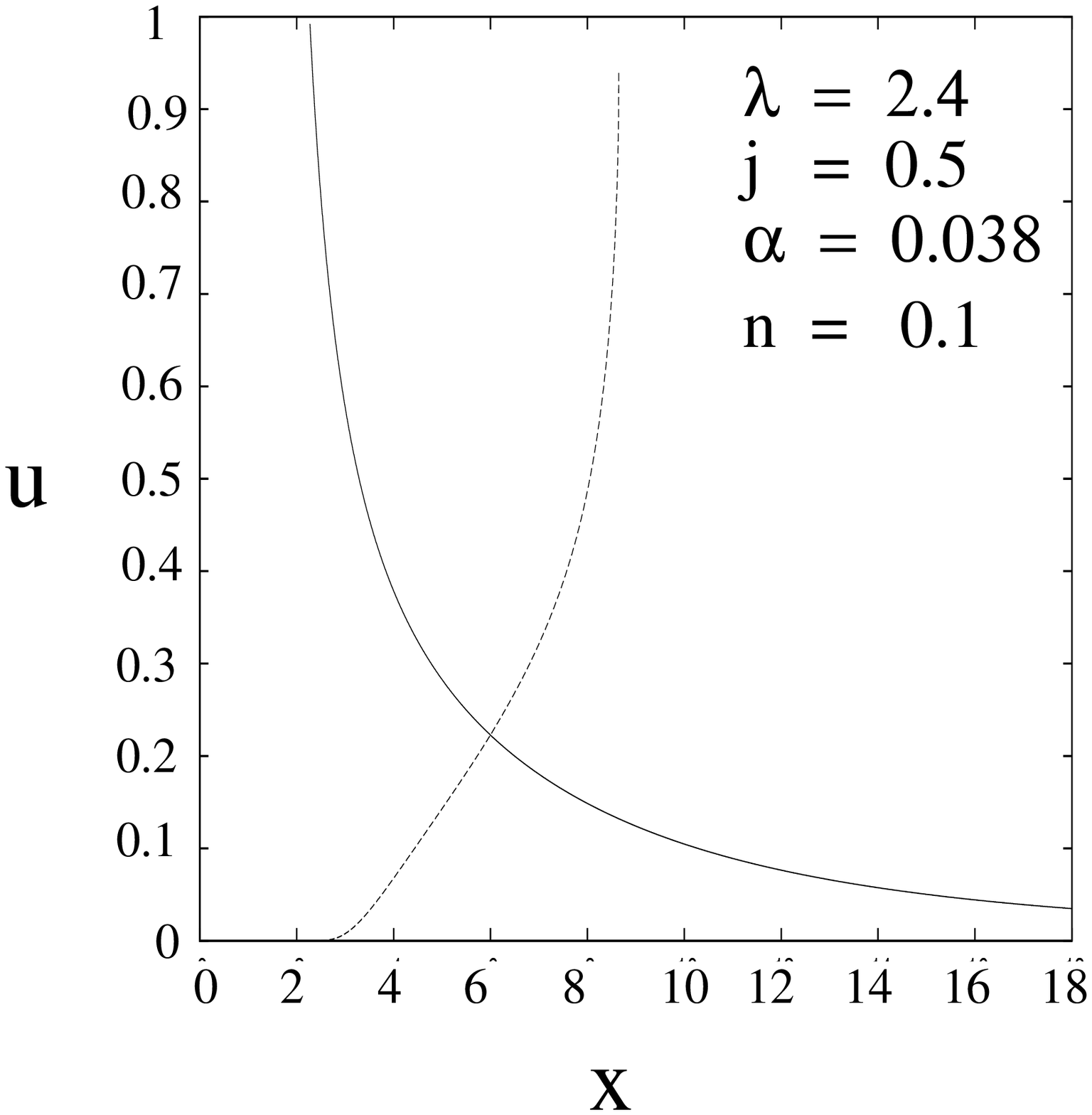}~~
\includegraphics[height=2.6in, width=2.6in]{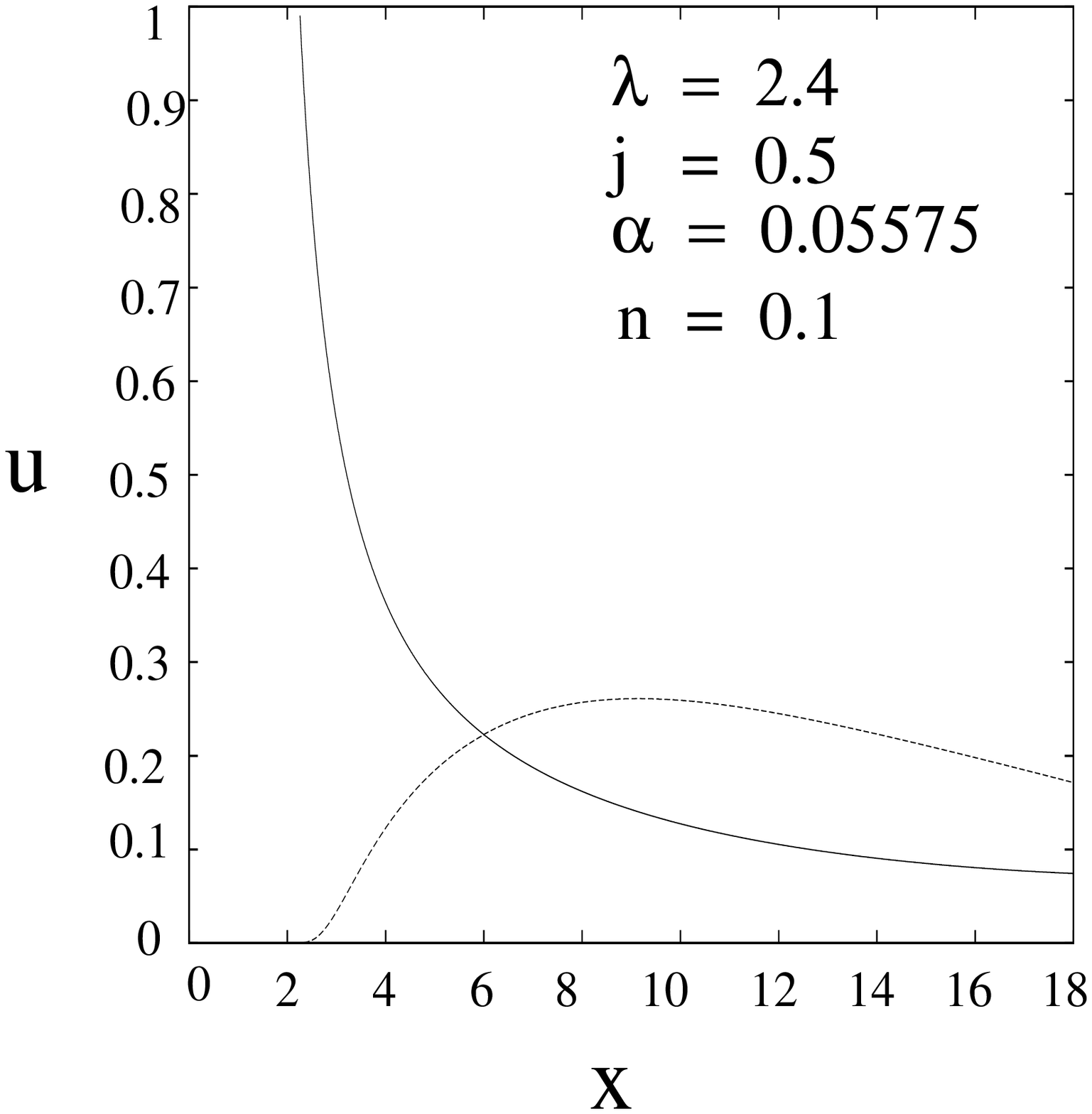}\\ \\
Fig. 3(a)-3(d) represent the variation of accretion and wind velocities in disc flows as functions of radial coordinate for $j=0.5$. The solid lines represent the accretion whereas the dotted lines are for wind. 
Note that the magnitude of velocities is plotted.
\hspace{1cm}

\vspace{2cm}

\end{figure}

\begin{figure}
(a)~~~~~~~~~~~~~~~~~~~~~~~~~~~~~~~~~~~~~~~~~~~~~~~~~~~~~~~~(b) \\ 
\includegraphics[height=2.6in, width=2.6in]{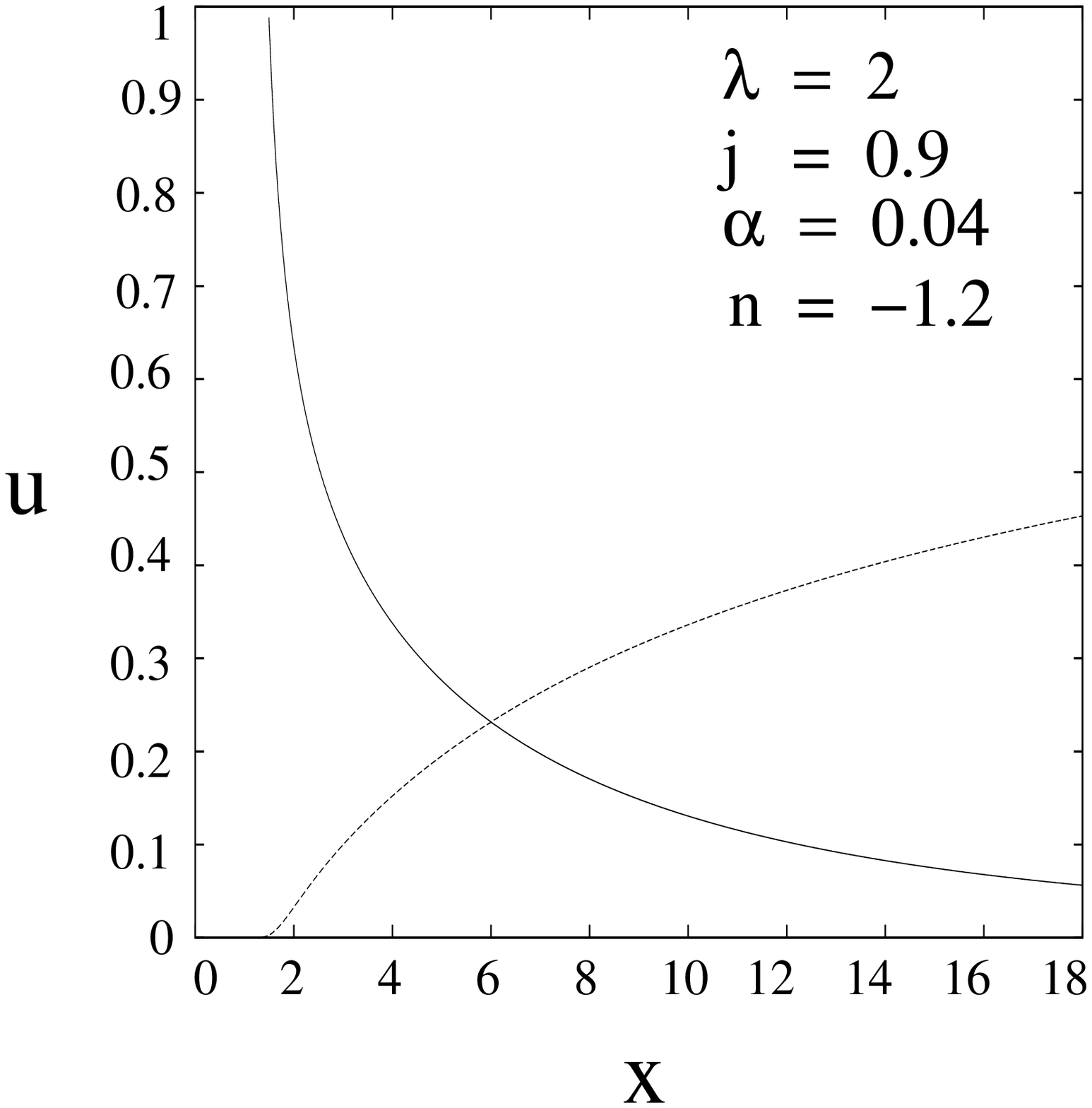}~~
\includegraphics[height=2.6in, width=2.6in]{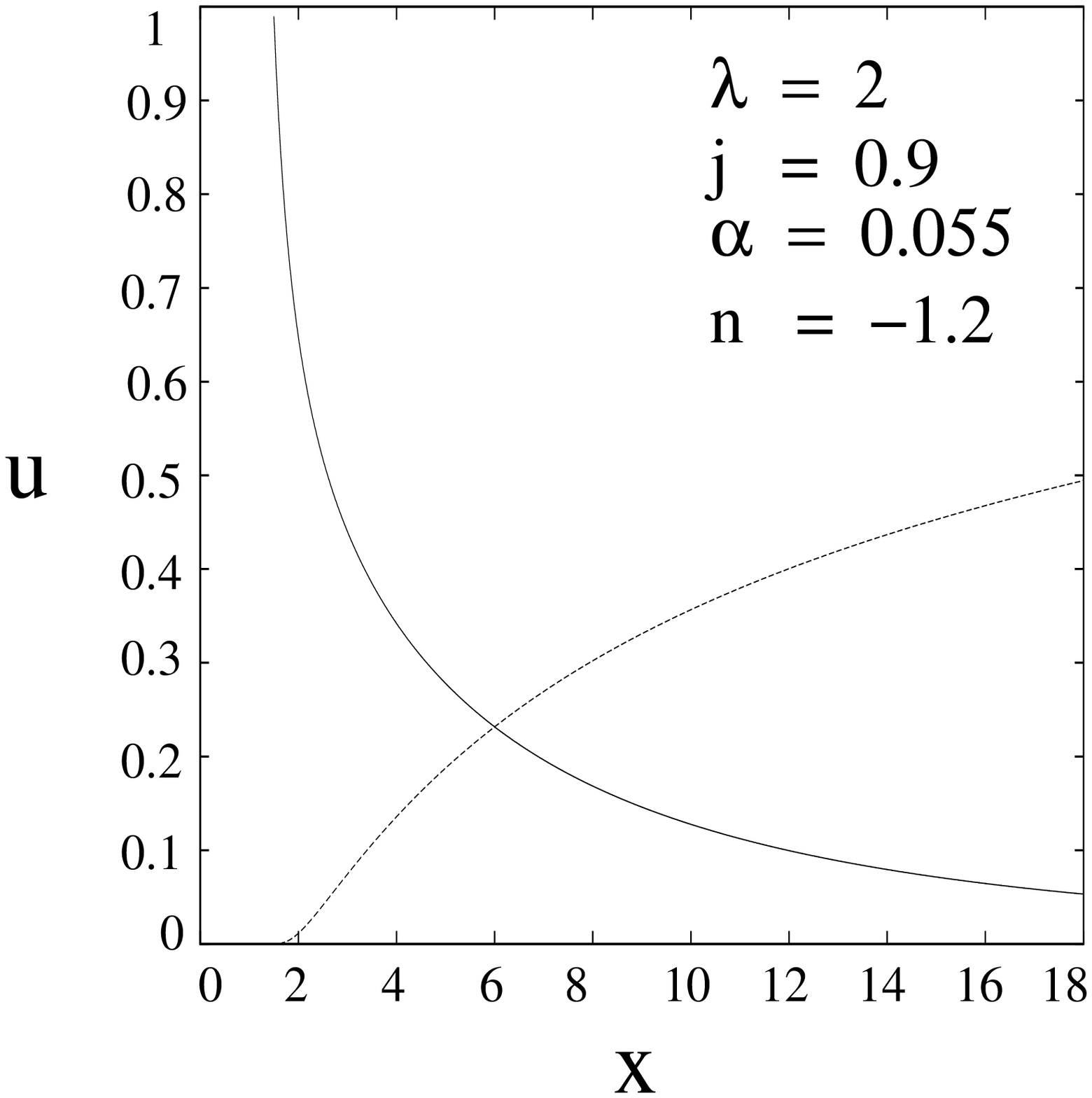}\\ \\
\hspace{1cm}
~~~~~~~~~~~~~~~~~~~~~~~~~~~~~~~~~~~~~~(c)~~~~~~~~~~~~~~~~~~~~~~~~~~~~~~~~~~~~~~~~~~~~~~~~~~~~~~~~(d) \\ 
\includegraphics[height=2.6in, width=2.6in]{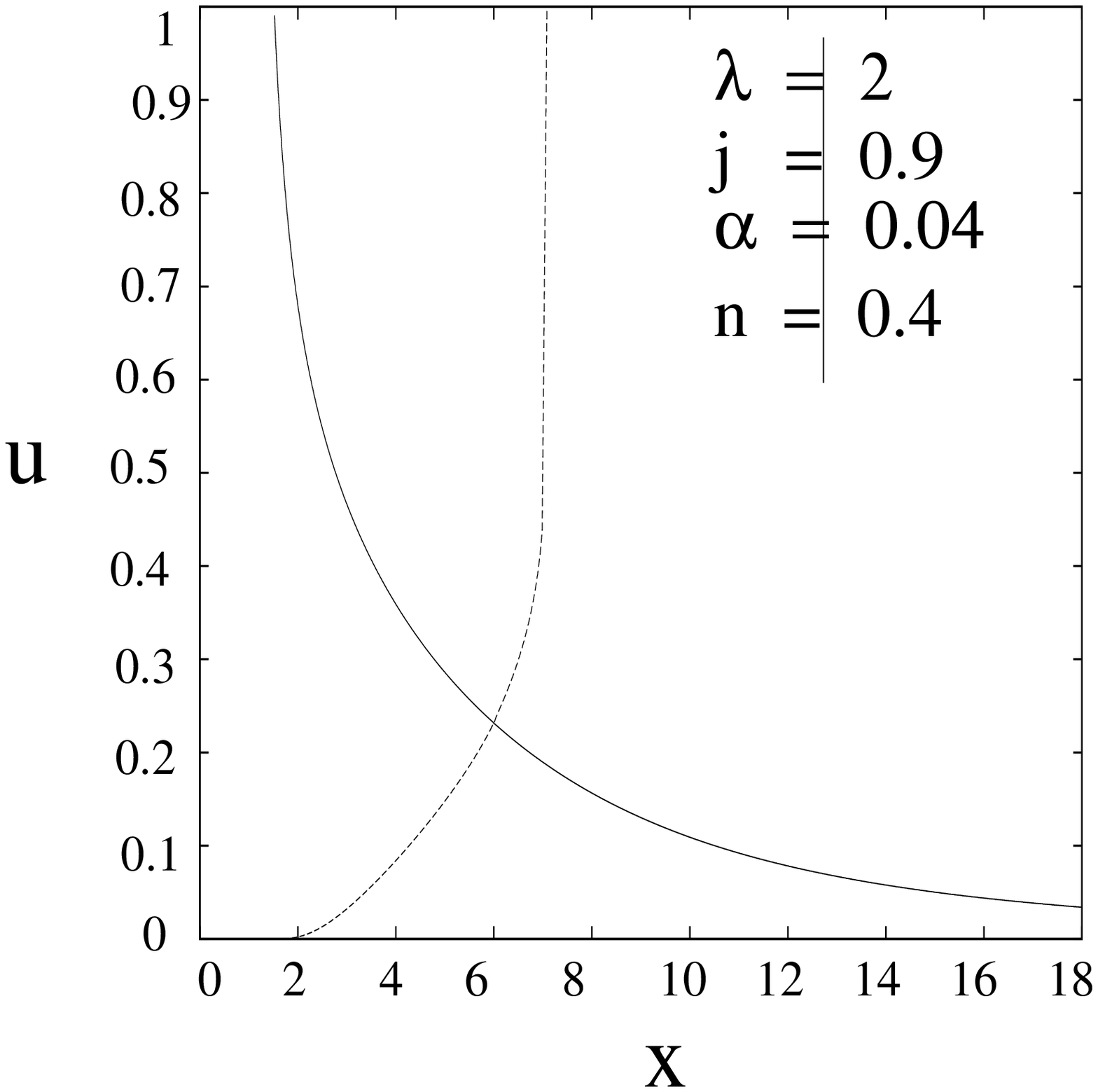}~~
\includegraphics[height=2.6in, width=2.6in]{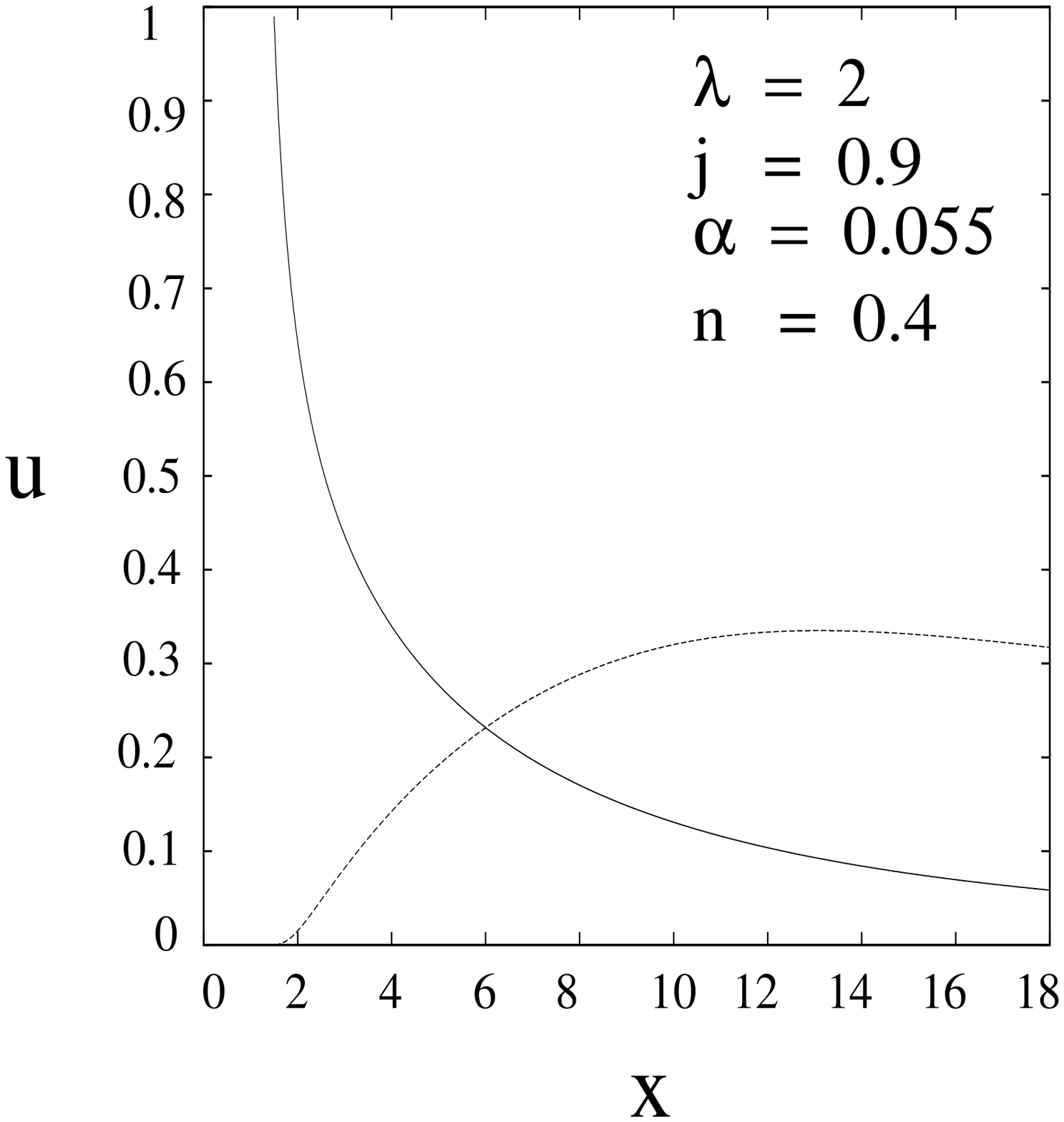}\\ \\
Fig. 4(a)-4(d) represent the variation of accretion and wind velocities in disc flows as functions of radial coordinate for $j=0.9$. The solid lines represent the accretion whereas the dotted lines are for wind. 
Note that the magnitude of velocities is plotted.
\hspace{1cm}
\vspace{2cm}

\end{figure}
\begin{figure}
(a)~~~~~~~~~~~~~~~~~~~~~~~~~~~~~~~~~~~~~~~~~~~~~~~~~~~~~~~~(b) \\ 
\includegraphics[height=2.6in, width=2.6in]{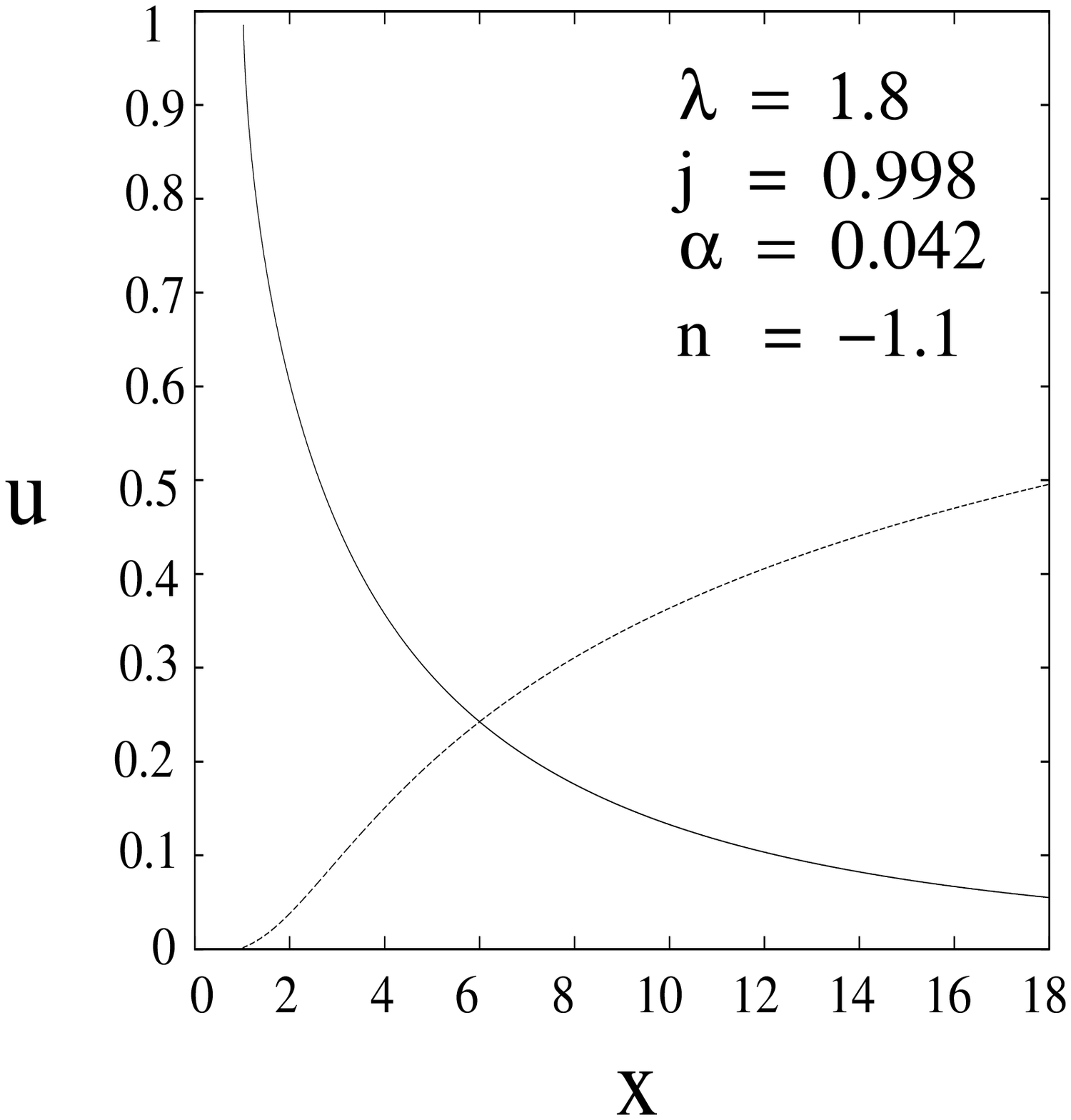}~~
\includegraphics[height=2.6in, width=2.6in]{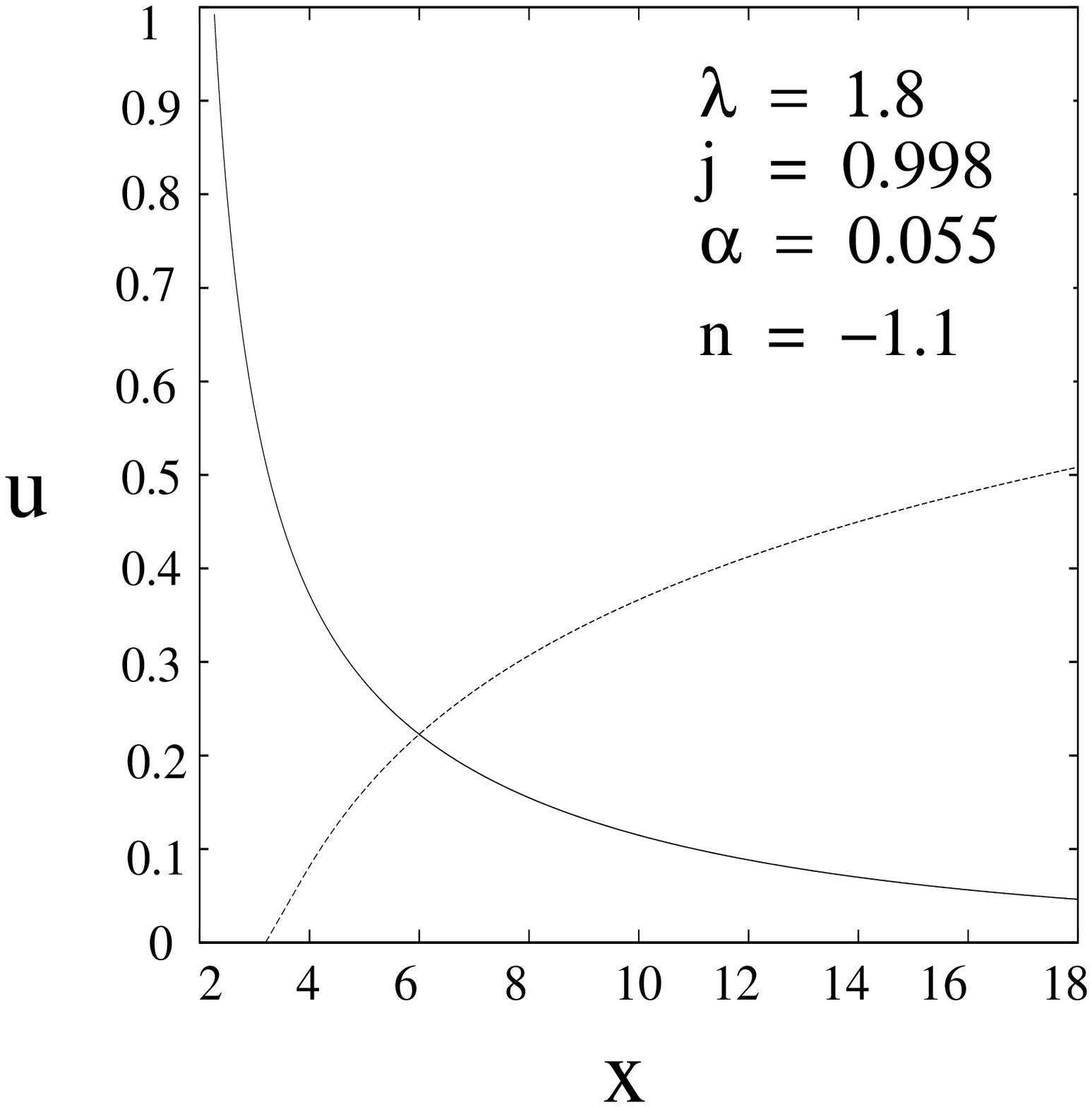}\\ \\
\hspace{1cm}
~~~~~~~~~~~~~~~~~~~~~~~~~~~~~~~~~~~~~~(c)~~~~~~~~~~~~~~~~~~~~~~~~~~~~~~~~~~~~~~~~~~~~~~~~~~~~~~~~(d) \\ 
\includegraphics[height=2.6in, width=2.6in]{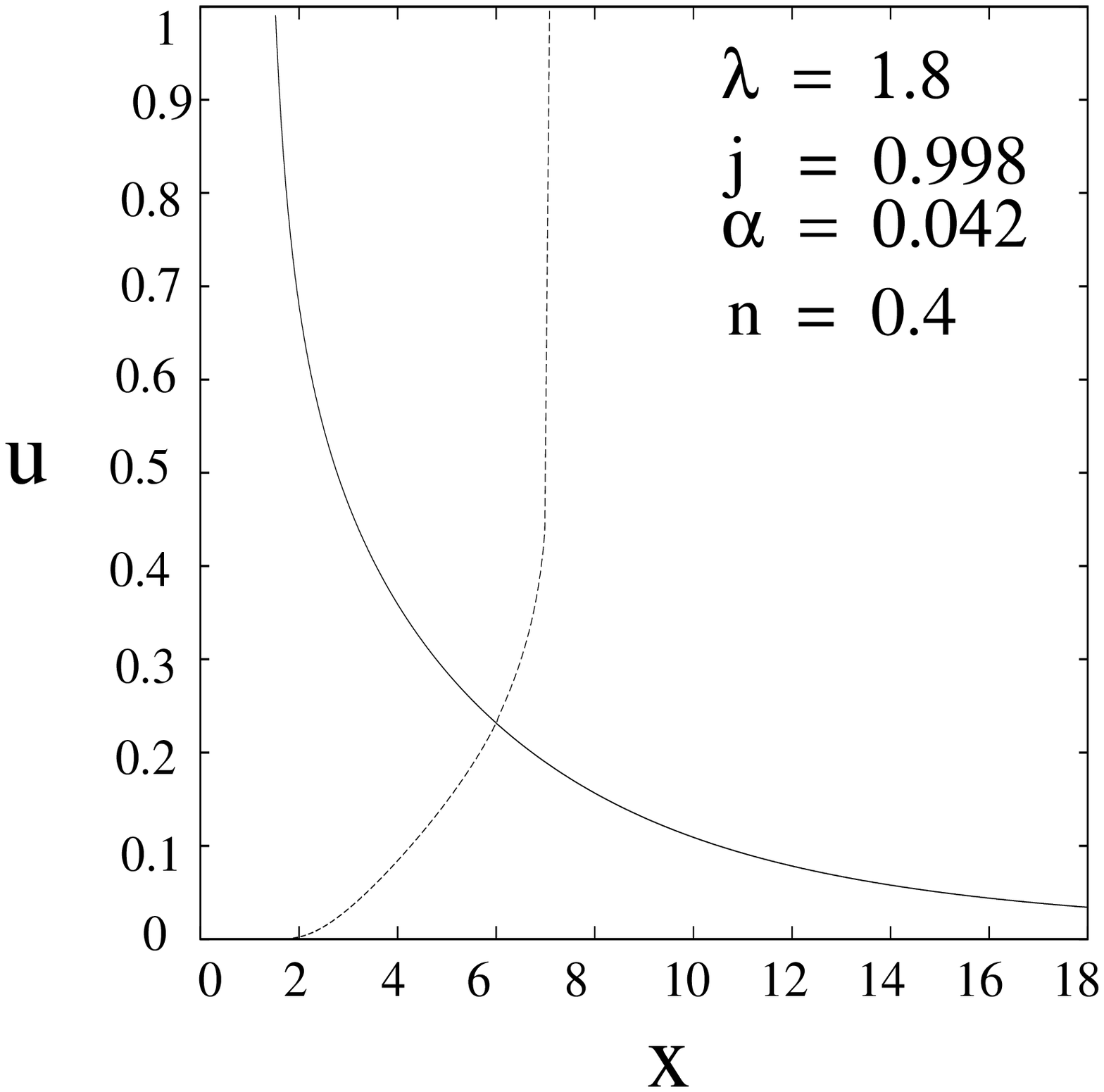}~~
\includegraphics[height=2.6in, width=2.6in]{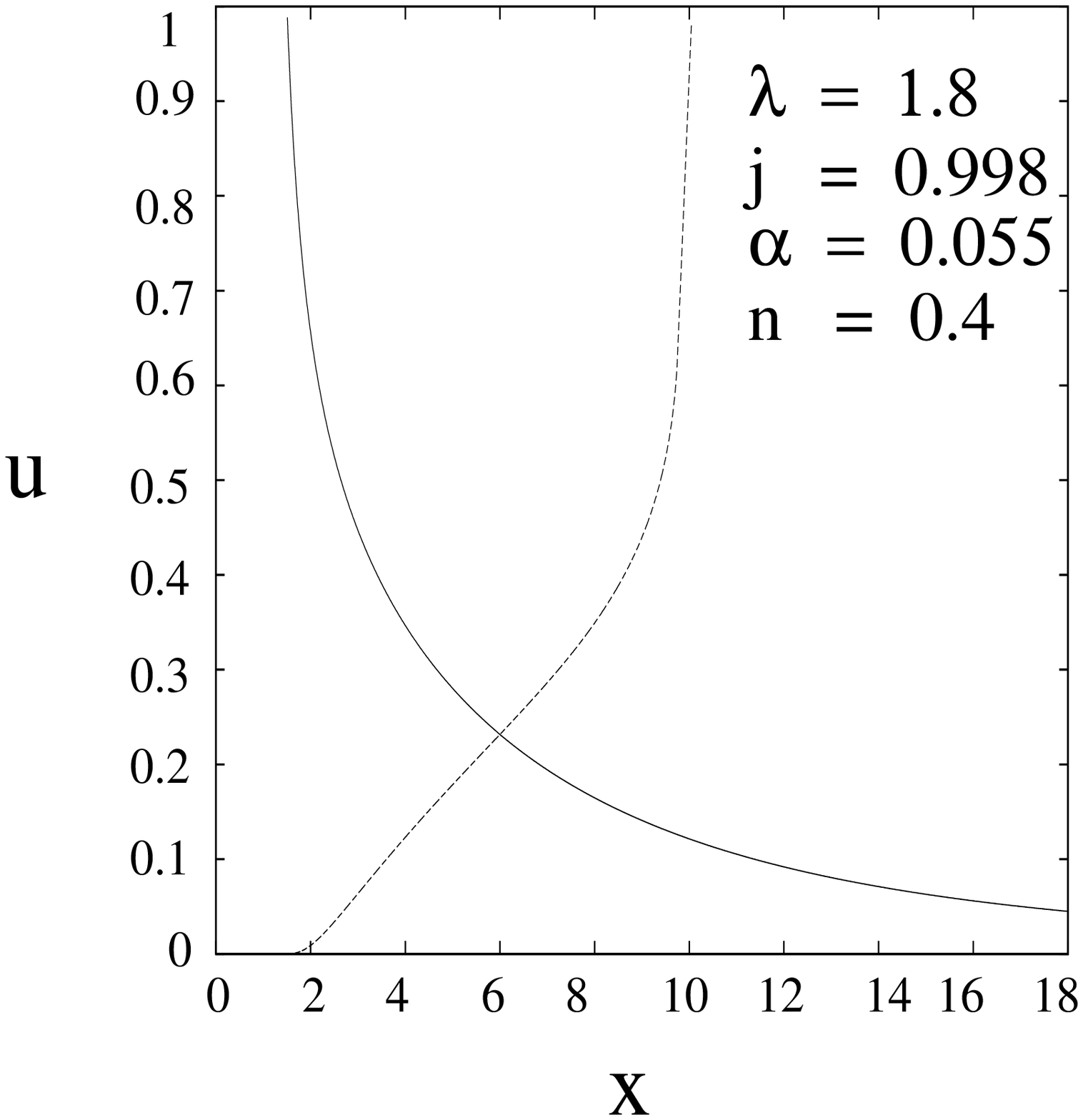}\\ \\
Fig. 5(a)-5(d) represent the variation of accretion and wind velocities in disc flows as functions of radial coordinate for $j=0.998$. The solid lines represent the accretion whereas the dotted lines are for wind. 
Note that the magnitude of velocities is plotted.
\hspace{1cm}

\vspace{2cm}

\end{figure}

Here we consider the DE accretion through spiral galaxies. Like Bondi solutions, two distinct curves
in figures described below represent the velocities of accreting matter and wind.
If the accretion disc has a higher specific angular momentum, then the corresponding centrifugal force tries to throw 
matter outwards. Therefore, wind increases which results in the shifting of critical point nearer to the BH.

As before the ranges of MCG parameters are fixed in order to obtain the physical solutions.
In figs 2(a)-2(d) we show the velocity profiles around a Schwarzschild BH with $\lambda=3.3$ 
for MCG. The critical point is at $x_{c}=4.45$ with $\alpha$ varying from $0$ to $0.07$ and $n$ from $-1.6$ to $0.6$ 
for a physical 
accretion-wind system. 
As MCG has a tendency to escape away and to increase the wind, matter has to come very near to the BH to
form the critical point, so 
that the stronger gravitational attraction therein can make accretion and wind velocities equal. 
However, when $\alpha=0$ and $-\frac{5}{3}<n<-\frac{4}{3}$, MCG is particularized into adiabatic gas for which the solutions obtained
by earlier work \cite{Mukhopadhyay1} are recovered, as shown in fig 2(a). We note that if $\lambda$ decreases, then the critical point shifts far from the BH. This is 
because, the decrease of rotational speed of accretion flow implies the decrease of centrifugal force. Hence, 
the wind velocity increases slowly
and becomes equal to the accretion velocity at a larger radius which results in forming the critical point away from the BH. 
However, an increase in `$\alpha$' keeping $n$ unchanged results in a shift of the critical point nearer to the BH.
Note that $\beta$, which is responsible for negative pressure, is constrainted by $\alpha$ and must be positive in order 
to have physical density. See \S 6 for further discussions in the issue.
This suggests that the MCG has a 
natural tendency of being thrown out from the BH system. On the other hand, higher $\alpha$ might render first term 
in equation (\ref{1}) dominating over the second, making MCG isothermal-like. This again results in a faster wind.
Therefore, as $\alpha$ increases, matter from accretion flow will be 
thrown outwards more rapidly. 
Further, as $n$ increases, the wind dominates with a very steeper velocity profile, as shown in fig 2(c), as 
discussed in order to explain fig 1(c). Moreover, with the increase of $\alpha$, as in fig 1(d), $\beta$ in fig 2(d) has to 
be negative rendering a slower wind.

In the rest of the figures we have considered the rotating BHs, i.e., the Kerr BHs. The figs 3(a)-3(d) are for 
Kerr parameter $j=0.5$, when the specific angular momentum of the accretion disc is fixed at $\lambda=2.4$ 
and the critical point $x_{c}=6$. 
Note that $\lambda$ decreases with the increase of $j$. This is because the higher gravitational force
acting radially at a high $j$ does not physically allow the disc to rotate faster and hence for a natural solution
disc centrifugal force and thus $\lambda$ has to decrease. 
When BH starts to rotate, its attractive power increases, and hence the critical point,
i.e., the radius where the accretion and wind velocities are same, forms away from the BH. The physically meaningful
ranges of values of $\alpha$ and $n$ turn out to be $0.038\leq\alpha\leq 0.05575$ and $-1.6\leq n\leq 0.1$. In fig 3(a),
$\alpha$ and $n$ are chosen to be very small and the solutions are very similar to that of the adiabatic ones for the Kerr BH as shown in
previous work \cite{Mukhopadhyay1}. Fig 3(c) shows that for a small $\alpha$ but large $n$ the wind velocity approaches the speed of light at a finite 
distance, close to the BH. This can be explained, recalling previous discussions for nonrotating BHs, as follows: 
increase of $n$ implies the abrupt increase of the negative term of the matter pressure
$-\frac{\beta}{\rho^{n}}$ for a very small $\rho$ of the DE dominated universe. Now negative pressure creates repulsion. Therefore, 
ultimately the increase of $n$ increases the repulsion in such a way that at a finite distance from the BH the tendency of matter
being thrown out increases and the matter velocity tends to become speed of light. As in the case of accretion around
nonrotating BHs, at a high $n$, once $\alpha$ changes to a highest (physically) possible value,
gas pressure becomes positive rendering a slower outflow.
In the realistic regime,
for minimum `$n$' and maximum `$\alpha$' the accretion-wind velocity profiles in fig 3(b) appear similar to that of the adiabatic ones. 
Low $n$ or negative $n$ implies (almost) adiabatic gas or (almost) isothermal gas equation of state, if $\rho$ is very small,
particularly for $\rho<1$: 
$p=\alpha \rho -\frac{\beta}{\rho^{n}}=\alpha \rho -\beta\rho^{N}\sim -\beta\rho^{N}\sim \beta_{2}\rho^{N}$ for $\beta_2>>\alpha$,
otherwise $\sim \alpha\rho$,
where $N=-n$ and $\beta_2=-\beta$. On the other hand, for large values of $n$ and $\alpha$ (within the admissible range), both the profiles of 
wind and accretion velocities exhibit decreasing slope far away from the BH (see fig. 3(d)) because of the resason explained
earlier.

Figs 4(a)-4(d) are for $j=0.9$ and $\lambda=2$. The admissible ranges for $\alpha$ and $n$ are 
$0.04\leq\alpha\leq 0.055$ and $-1.2\leq n\leq 0.4$ and the critical point is at $x_{c}=6$. These figures are quite similar to 
figs 3(a)-3(d).

For a rapidly rotating BH with $j=0.998$, the critical point is fixed at $x_{c}=6$ and $\lambda$ is chosen to be $1.8$. 
The possible ranges for `$\alpha$' and `$n$' are $0.042\leq\alpha\leq 0.055$ and $-1.1\leq n\leq 0.4$. The velocity profiles 
are presented in figs 5(a)-5(d). In figs 5(a) and 5(b), it is shown that for small `$n$' the solutions are very similar to 
that in the adiabatic cases irrespective of the choice of $\alpha$. In figs 5(c)-5(d), it has 
been shown that for large `$n$' wind velocity increases abruptly and tends to become the speed of light close
to the BH, at a finite 
distance from the critical point, as in the cases of Bondi and some of the disc flows due to dominance of negative pressure, 
while the accretion flow has the same nature as in figs 5(a)-(b).

\section{Accretion-wind solutions for viable MCG parameters: Constraints from the current observed data}
In \S 3 and \S 4, in drawing the accretion and wind curves we have emphasized the changes in the solutions 
for different extreme values of $\alpha$ and $n$. We have essentially considered the ranges of parameters
which give rise to the physical solutions, i.e. the solutions extending from infinity to the black hole horizon.
In order to obtain such solutions, even we have chosen $n<0$, which does not support the observed data. 
The main motivation to choose $n$ negative 
is to recover the results obtained for adiabatic gas\footnote{The choice of $\alpha=0$ and 
$-\frac{5}{3}<n<-\frac{4}{3}$ renders MCG to adiabatic gas.} in the present framework.

Based on dimensionless age parameter ($H_{0}-t_{0}$) \cite{Dev1} and observed $H(z)-z$ \cite{Wu1} data for both 
cold dark matter (CDM) and unified dark matter energy (UDME) models, the values of parameters $\alpha$ and 
$n$ are constrainted \cite{Thakur1}. 
In order to obtain a viable cosmology with MCG, $\alpha$ should be restricted to positive values. 
Besides this, there are best-fit values of the parameters for CDM and UDME models which correspond to
$H_0-t_0$ data. The permissible values of the parameters are given in the following table.

\begin{center}
{\bf Table I:} Permissible ranges of MCG parameters in different models.\\
\begin{tabular}{|l|}
\hline\hline
~~Source of data
~~~~~~~~~~~~~~~~~~~~~~~~~~~~~~~~~~~~~~~~~~~~~~~~~~~~~~~~Type of Model \\ \hline
\\
~~~~~~~~~~~~~~~~~~~~~~~~~~~~~~~~~~~~~~~~~~~~~~~~~~~~~~~~~~~~~~CDM~~~~~~~~~~~~~~~~~~~~~~~~~~~~~~~~UDME~~~~~~~
\\\hline
~~~~~~~~~~~~~~~~~~~~~~~~~~~~~~~~~~~~~~~~~~~~~~~~~~~~~~$\alpha$~~~~~~~~~~~~~~~~~~$n$~~~~~~~~~~~~~~~~~~~~$\alpha$~~~~~~~~~~~~~~~~~~$n$~~~~~~~
\\\hline\hline
\\
~$H(z)-z$~~~~~~~~~~~~~~~~~~~~~~~~~~~~~~~~~$0\leq\alpha\leq 1.07$~~~~~~$0\leq n\leq 1$~~~~~~~$0\leq\alpha\leq 1.35$~~~~~$0\leq n\leq 1$
\\\\\hline\\
~$H_{0}-t_{0}$~~~~~~~~~~~~~~~~~~~~~~~~~~~~~~~~~~~~~~$\alpha=0.01$~~~~~~~~~$ n=0.01$~~~~~~~~~~$\alpha=0.06$~~~~~~~~~$ n=0.11$~\\
(best-fit)
\\\\ \hline\hline
\end{tabular}
\end{center}

In \S\S 3 and 4, all the solutions, except the ones with negative $n$ (reproducing adiabatic cases, particularly
for $\alpha=0$), are in accordance with observed constraint on $\alpha$ and $n$.
In figs 6(a)-6(d) we show the accretion-wind solutions for best-fit $\alpha$ and $n$ given in Table I. 
We consider different combinations of $\lambda$ and $j$. Fig 6(a) is for spherical accretion, i.e. 
$\lambda=j=0$. As $\alpha=n=0.01$, for $\rho<1$ (which is the case in our solutions) the magnitude of first term 
in equation (\ref{1}) ($\alpha\rho$) is negligible compared to that of second term ($\beta\rho^{-n}$). 
Hence, flow will have negative pressure, but less negative compared to that in fig 1(c) which is for
$40\%$ higher $n$. Therefore, 
in absence of centrifugal barrier the critical point forms away off the BH compared to that in fig 1(c). 
Here we see that no proper wind solution is found for $x_{c}\le 11.55$. 
Therefore, the solutions are for $x_c=11.55$. 

\begin{figure}
(a)~~~~~~~~~~~~~~~~~~~~~~~~~~~~~~~~~~~~~~~~~~(b) \\
\includegraphics[height=2.6in, width=2.6in]{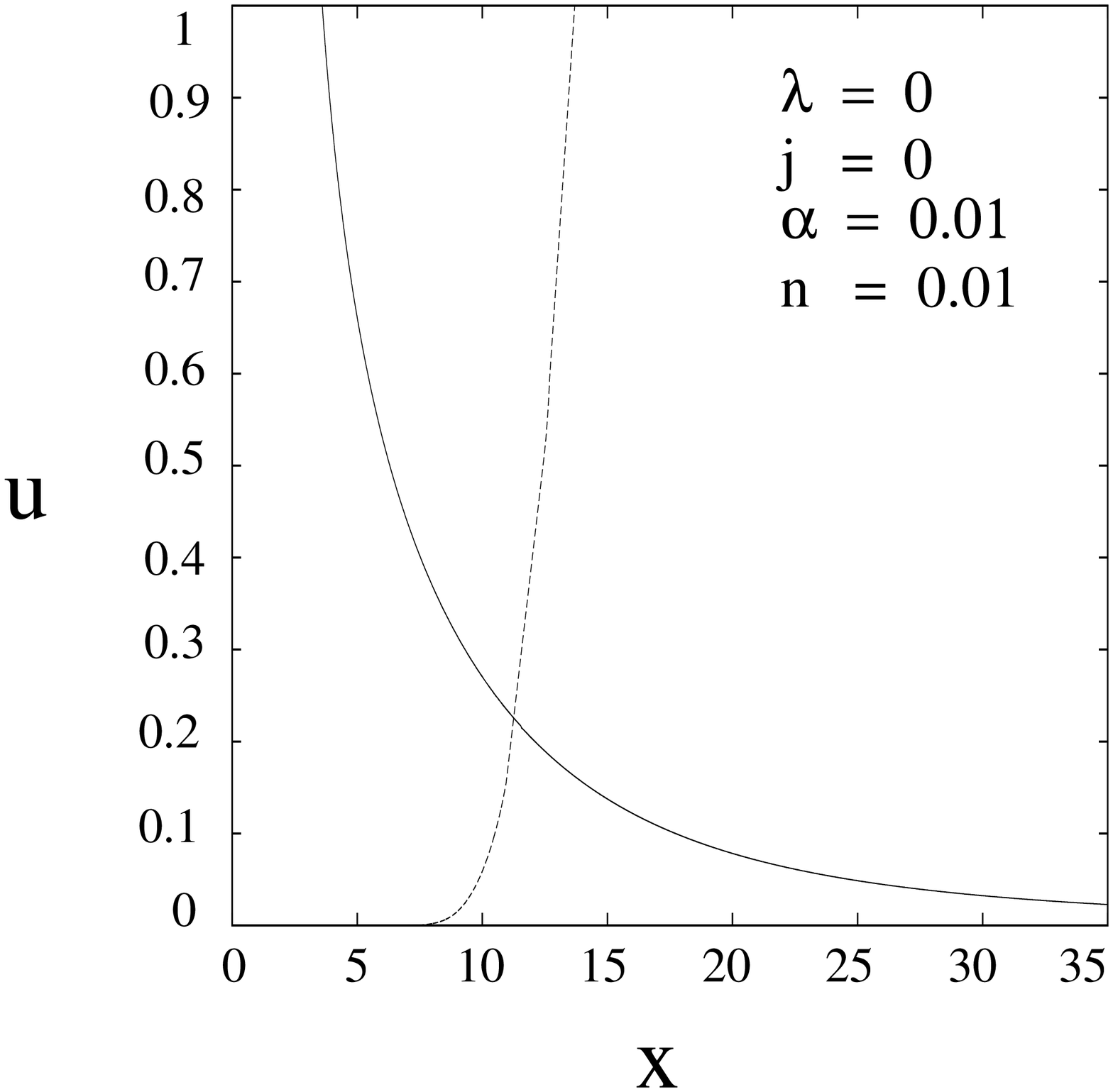}~~
\includegraphics[height=2.6in, width=2.6in]{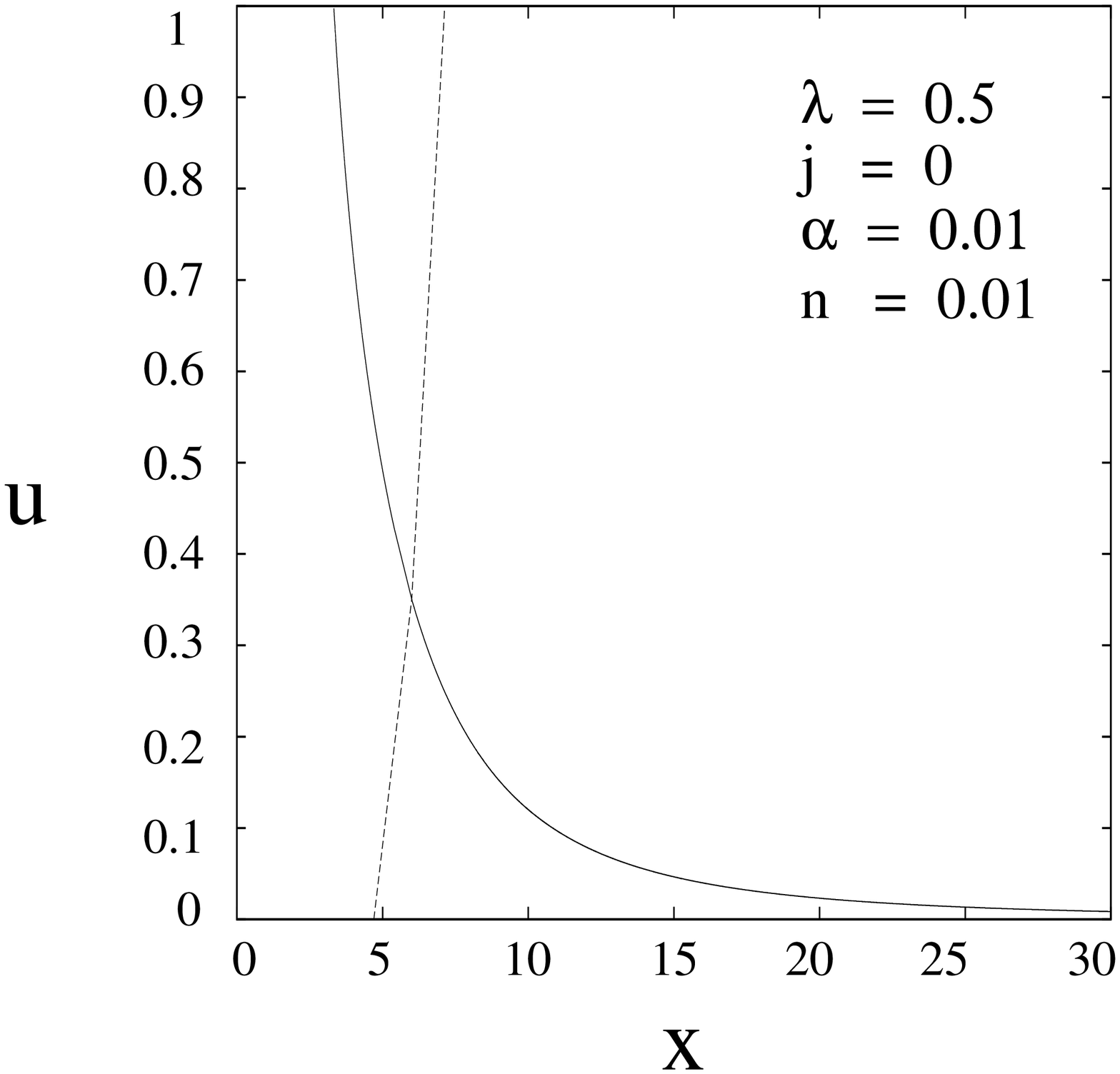}\\\\
\hspace{1cm}
(c)~~~~~~~~~~~~~~~~~~~~~~~~~~~~~~~~~~~~~~~~~~(d) \\
\includegraphics[height=2.6in, width=2.6in]{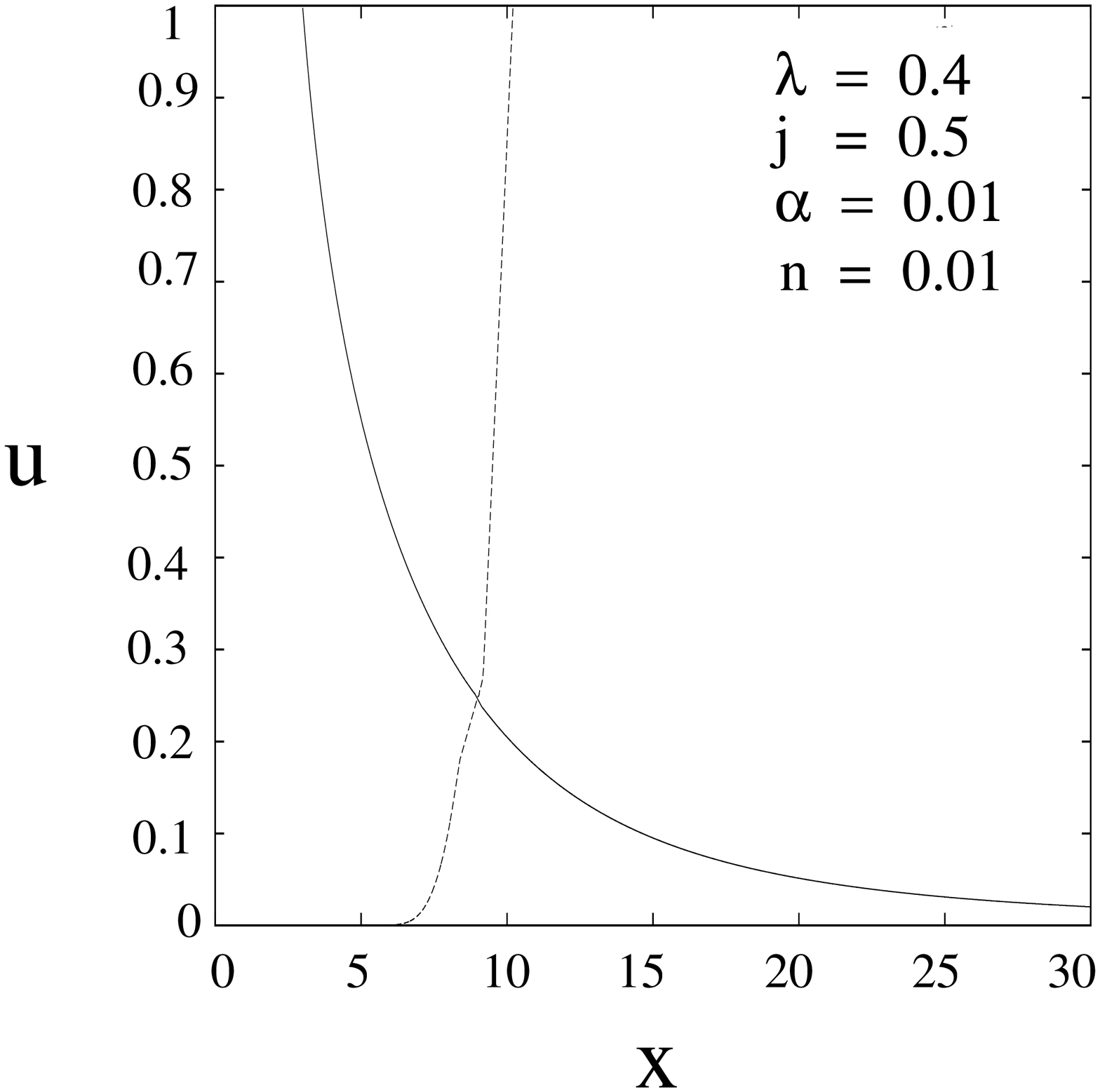}~~
\includegraphics[height=2.6in, width=2.6in]{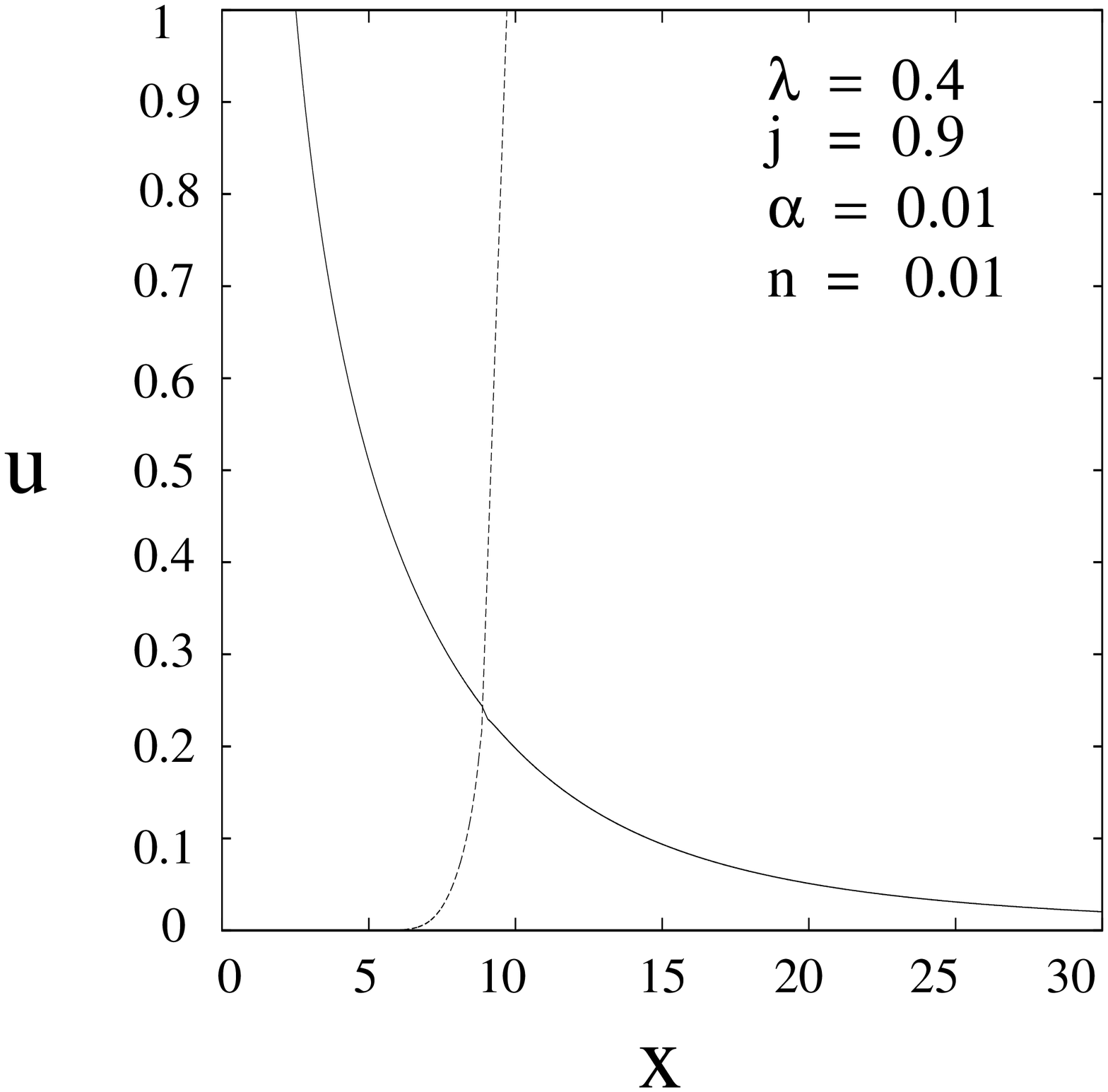}\\ \\
Fig. 6(a)-6(d) represent the variation of accretion and wind velocities in disc flows as functions of radial 
coordinate with $\alpha=0.01$ and $n=0.01$ (best-fit parameters), for (a) Bondi flow, (b), (c), (d) for 
disc flows around the BH with $j=0,0.5,0.9$ respectively. The solid lines represent the accretion
whereas the dotted lines are for wind.
Note that the magnitude of velocities is plotted.
\hspace{1cm}
\vspace{2cm}

\end{figure}

Figure 6(b) is for disc accretion around a Schwarzschild BH. Due to presence centrifugal barrier radial
velocity of matter decreases away from the BH compared to that in fig 6(a) and hence the critical point 
forms at an inner radius. We choose $x_{c}=6$ for fig 6(b). Although the nature of accretion branch is 
similar to that in the figure 6(a), the accretion velocity is very small far from the BH. On the other
hand, slightly away from the critical point the wind velocity increases steeply and becomes almost equal to the 
speed of light at a smaller radius than that in the Bondi flow shown in fig 6(a). This is because
the centrifugal force counteracts the gravitational force until very inner region of the disc. 
Increase of $\lambda$ would make the wind even stronger having velocity close to speed of light 
at a radius even nearer to the BH. However, compared to the wind profile in presence of a stronger negative
pressure (e.g. that in fig 2(c)) counteracting gravitational force, the steepness of wind velocity profile remains 
small and hence the maximum possible $\lambda$ is restricted to a smaller value. 

Figure 6(c) is for disc solutions around a rotating BH with $j=0.5$ and maximum possible $\lambda$. 
Increase of $j$ increases the gravitational power of the BH, resulting in a shift of the critical point 
away from the BH compared to that in fig 6(b) and forms at $x_{c}=9$ 
Finally, solutions in fig 6(d) is for $j=0.9$ with $x_{c}=9$ and $\lambda=0.4$, which are very 
similar to that in 6(c).

\section{Variation of sound speed in accretion-wind flows}

\begin{figure}
(a)~~~~~~~~~~~~~~~~~~~~~~~~~~~~~~~~~~~~~~~~~~(b) \\
\includegraphics[height=2.6in, width=2.6in]{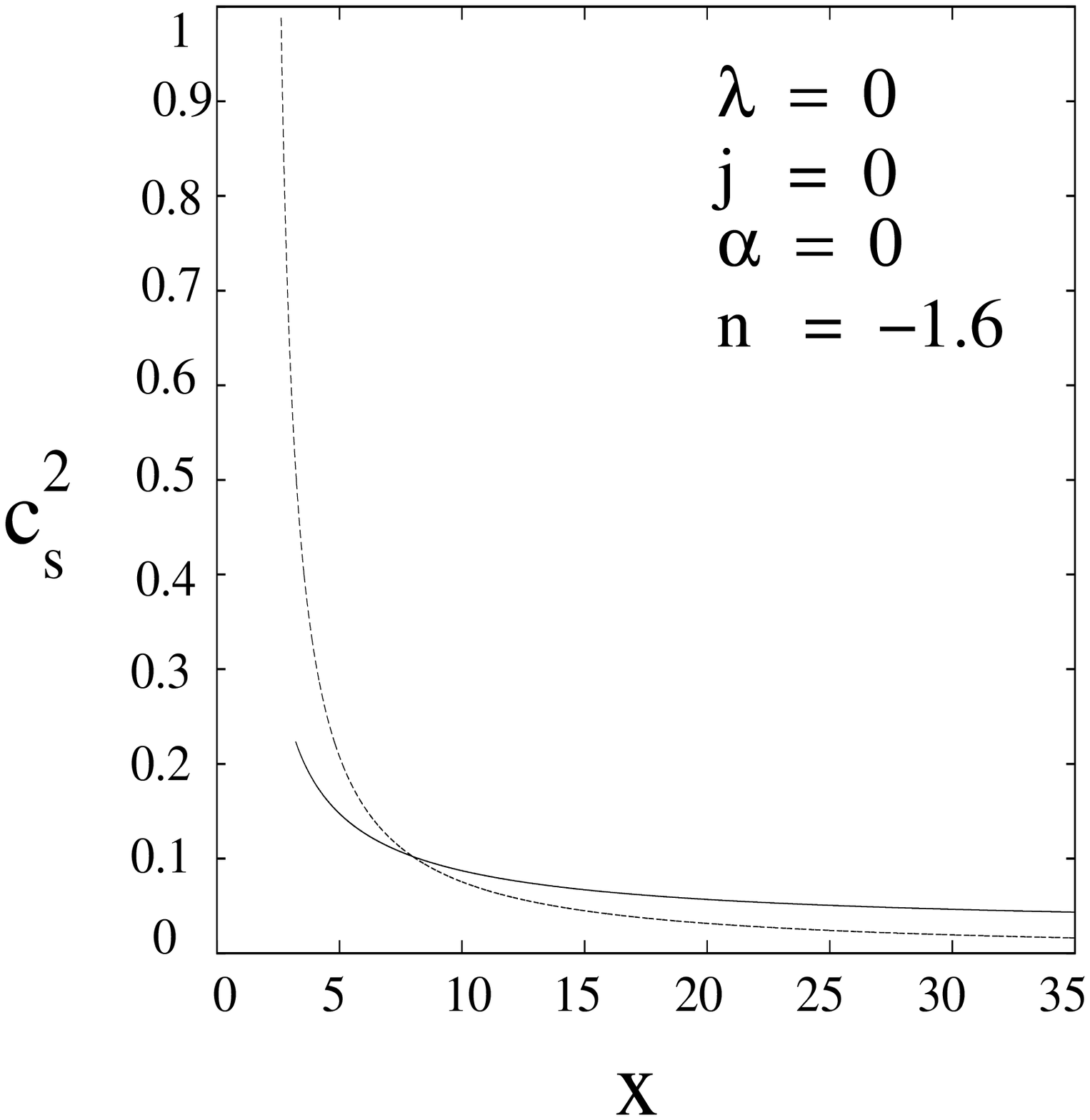}~~
\includegraphics[height=2.6in, width=2.6in]{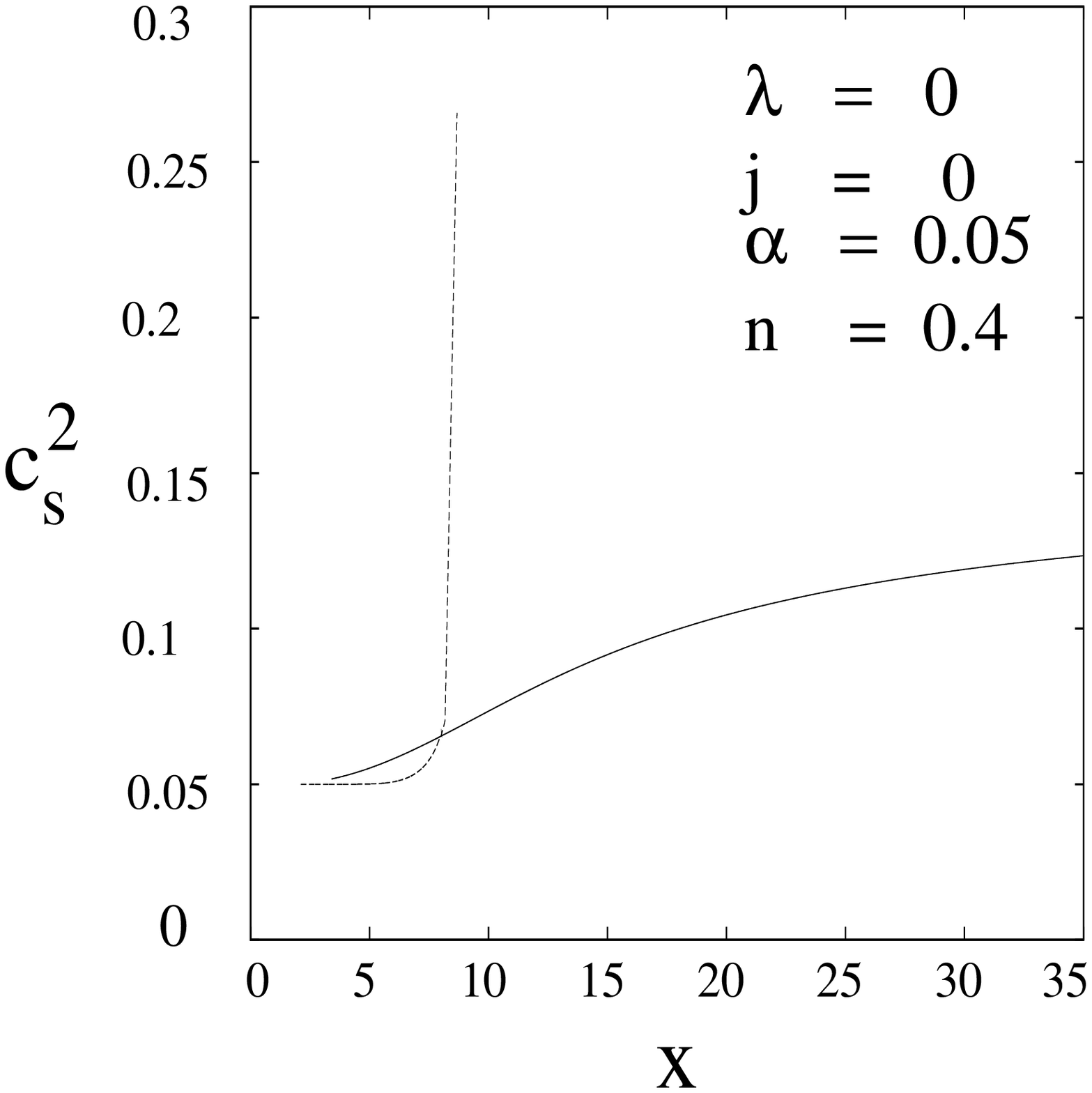}\\\\
\hspace{1cm}
(c)~~~~~~~~~~~~~~~~~~~~~~~~~~~~~~~~~~~~~~~~~~(d) \\
\includegraphics[height=2.6in, width=2.6in]{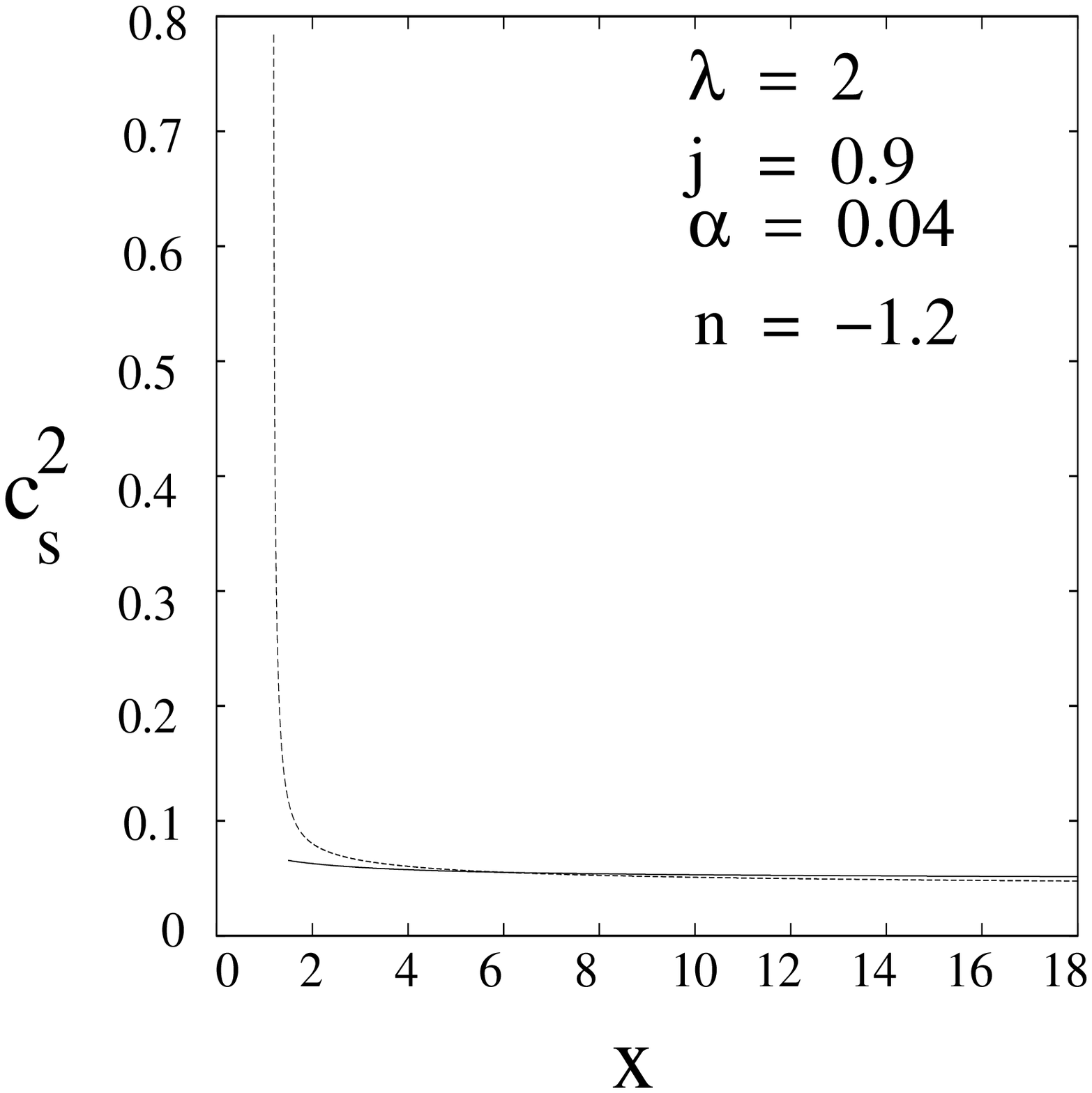}~~
\includegraphics[height=2.6in, width=2.6in]{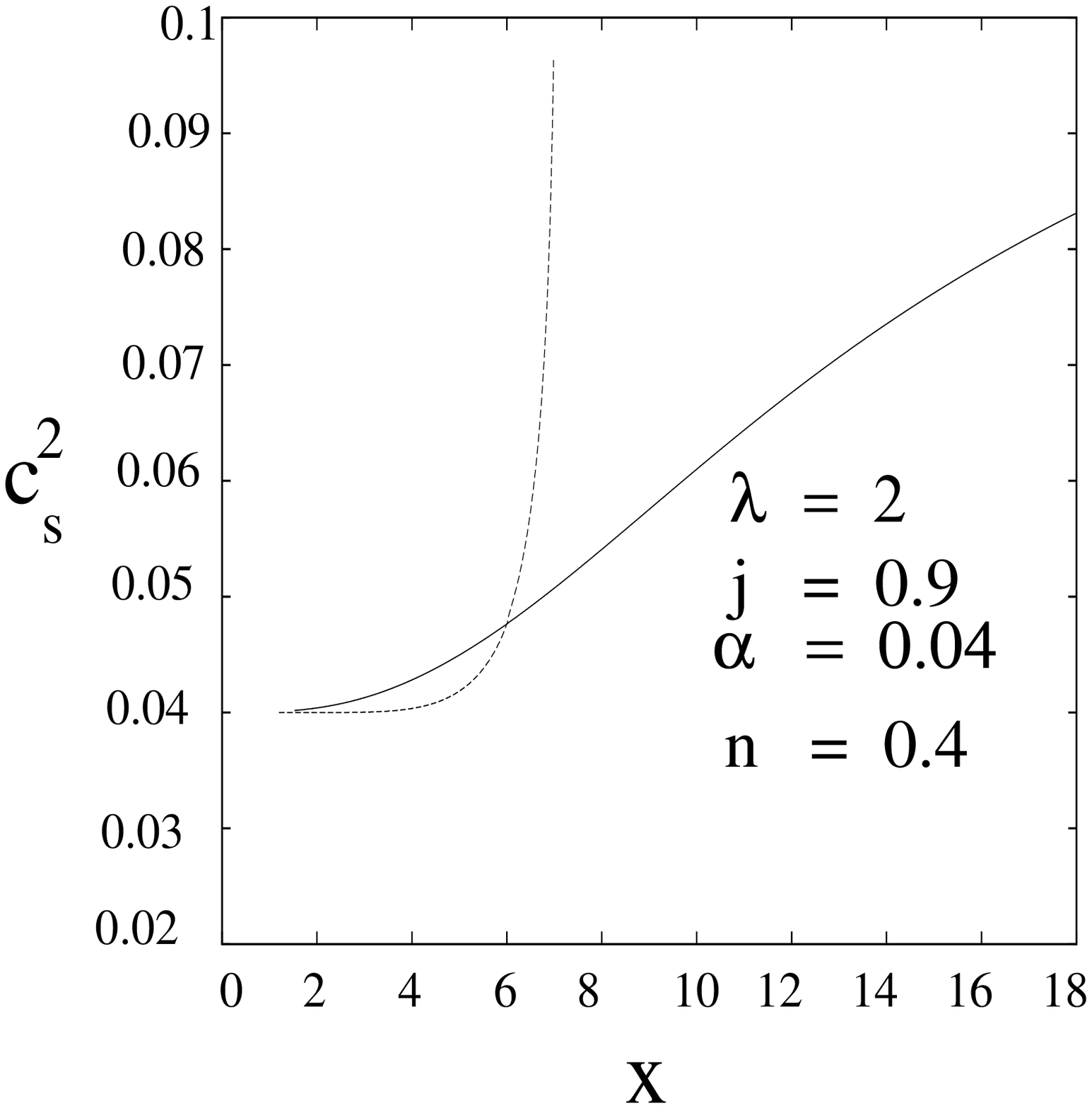}\\\\
Fig. 7(a)-7(d) represent the variation of square of sound speeds in accretion and wind flows as functions of 
radial coordinate corresponding to the velocity profiles shown in (a) Fig. 1(a), (b) Fig. 1(c),
(c) Fig. 4(a), (d) Fig. 4(d). The solid lines represent the accretion
whereas the dotted lines are for wind.
\hspace{1cm}
\vspace{2cm}
\end{figure}

Equation (\ref{6}) implies that $c_{s}^{2}>\alpha$, provided $\beta, n>0$. However, only positive $\beta$ 
assures negative pressure when $\alpha\ge 0$ (which is constrainted from observed data). Any solution 
with $c_{s}^{2}<\alpha$ does not imply MCG flow. Therefore, in realistic MCG flows, $c_{s}$ must have
a lower bound which is $\sqrt{\alpha}$. Below we discuss a few typical sound speed profiles in the flows
described above. 

Figures 7(a) and (c) represent the variation of sound speed
square in typical Bondi and disc accretion flows of adiabatic/adiabatic-like/isothermal-like gas. On the other hand,
figs 7(b) and (d) represent same for MCG showing clear lower bounds as respective values of $\alpha$. 
It is clear from equation (\ref{ex1}) that $c_s$ decreases with the increase of $\rho$ for MCG. Hence
$c_s^2$ in figs 7(b) and (d)
decreases as a function of radial coordinate, unlike that in flows shown in figs 7(a) and (c).
Note that $\rho$ always increases as matter approaches the BH.
If $c_{s}^{2}\rightarrow\alpha$, then $\rho\rightarrow\infty$. This is only possible when flow is very 
close to the BH event horizon, which indeed is the case in figs 7(b) and 7(d) when $c_{s}^{2}\rightarrow\alpha$ 
at $r\rightarrow$ event horizon.

Figure 7(a) represents a Bondi flow corresponding to the case shown in fig 1(a). Here the square of sound speed along the wind branch is less than that in 
accretion branch when $x>x_{c}$. As $x<x_{c}$ and matter approaches nearer to the BH, $c_{s}^{2}$ in wind increases 
rapidly, whereas $c_{s}^{2}$ in accretion increases much slowly. 
Figure 7(b) corresponds to the flow in fig 1(c). Here both the accretion and wind sound speeds are very high far away
from $x_{c}$ compared to that in the vicinity of the BH. However, the very high sound speed in wind in fig 7(b) 
resembles that of close to the BH in fig 7(a). Similar features appear in figs 7(c) and 7(d) where the sound speed,
corresponding to figs 4(a) and 4(c) showing disc flows around rotating black holes, is shown.

%

\section{Summary}

We have studied the accretion-wind solutions for MCG around BH. As the universe is expanding and is supposed, in many models, to be filled up with 
matter having negative pressure, it is very interesting to understand what happens to the BH accretion and wind when infalling matter is MCG. 
By nature, MCG has a positive pressure at early stage of the universe and then the pressure becomes negative at latter epochs. 
Therefore, at late stages the universe experiences an accelerated expansion and as a result normal accretion procedure must be impacted 
deeply by the infall of this matter. Indeed our results show that DE in the form of MCG is prone to be thrown outwards from the accretion flows. 
As the negative pressure increases with the increase of $n$, the wind velocity approaches to the speed of light at a near vicinity of the 
the BH when $\rho<1$, which is generically true for flows around BHs. 
In general the accretion-wind system of MCG dramatically alters the wind solutions, producing
faster winds, upon changes
in physical parameters, while accretion solutions qualitatively remain unaffected. This is, however, a natural consequence of the DE into the system.

Note importantly that our velocity profiles do not depend on $\dot{M}$ explicitly due to the inviscid
assumption when there is no energy equation to be considered. However, it
is of considerable interest to estimate the actual amount of MCG accretion
onto the super-massive black holes to address the issue of observational
consequences of our results which could explain the observed rapid growth of black holes at high
redshift \cite{fan}. Babichev et al. \cite{Babichev1} considered the scenario
where DE is accreted from the inter-galactic medium. Therefore, the accretion
rate in their work is determined by the cosmological evolution of the averaged
DE density in the universe. In our view, since the BHs we consider are
siting already in the gravitational potential of galaxies (not in that sense isolated BHs), 
we expect that even before BHs start accreting, host galaxy
would have already accreted some amount of DE. Hence the actual rate of
accretion of DE onto BHs should really be computed from the ambient value of
DE density inside host galaxies. We shall consider a detailed analysis
of this problem/issue in a later publication.


\begin{contribution}

{\bf Acknowledgement :}

RB thanks West Bengal State Government for awarding JRF, then 
IUCAA, Pune, as a part of this work was done during a visit and Mr. Aditya Kumar Naskar for several technical helps. SC is thankful to DST-PURSE Programme, Jadavpur University. 
BM acknowledges the support provided by a project (Grant No. SR/S2/HEP12/2007)
funded by Department of Science and Technology (DST), India. 

\end{contribution}

\end{document}